\documentclass[longauth]{aa}

\usepackage[toc,page]{appendix}
\usepackage{gensymb}
\usepackage{graphicx}
\usepackage{natbib}
\usepackage{scalerel}
\usepackage{adjustbox}
\usepackage{ulem}

\usepackage[table]{xcolor}

\usepackage[nolist,nohyperlinks]{acronym}

\usepackage{txfonts}
\usepackage[pdfencoding=auto,psdextra]{hyperref}
\hypersetup{
    colorlinks=true,
    linkcolor=blue,
    filecolor=magenta,      
    urlcolor=blue,
    citecolor=blue
}
\usepackage{orcidlink} 
\newcommand{\orcid}[1]{\orcidlink{#1}}

\usepackage[nameinlink,capitalise]{cleveref}
\crefname{section}{Sect.}{Sects.}
\Crefname{section}{Section}{Sections}
\crefname{figure}{Fig.}{Figs.}
\Crefname{figure}{Figure}{Figures}
\crefname{equation}{Eq.}{Eqs.}
\Crefname{equation}{Equation}{Equations}

\makeatletter
\renewcommand*\aa@pageof{, page \thepage{} of \pageref*{LastPage}}
\makeatother
\usepackage{lastpage}

\usepackage[utf8]{inputenc}

\usepackage[switch, modulo]{lineno}
\nolinenumbers

\usepackage{euclid}
\begin{document} 

\acrodef{AA}{$\alpha$ angle}
\acrodef{ADS}{Airbus Defence and Space}
\acrodef{BFE}{brighter-fatter effect}
\acrodef{BOL}{beginning of life}
\acrodef{CCD}{charge-coupled device}
\acrodef{CDR}{Critical Design Review}
\acrodef{CPPM}{Centre de Physique des Particules de Marseille}
\acrodef{CR}{cosmic ray}
\acrodef{CSL}{Centre Spatial de Li\`ege}
\acrodef{CTE}{coefficient of thermal expansion}
\acrodef{DPU}{Data Processing Unit}
\acrodef{DS}{Detector System}
\acrodef{EDS}{\Euclid Deep Survey}
\acrodef{EOL}{end of life}
\acrodef{ERO}{Early Release Observations}
\acrodef{ESA}{European Space Agency}
\acrodef{EWS}{\Euclid wide survey}
\acrodef{FAR}{Flight Acceptance Review}
\acrodef{FEM}{finite element model}
\acrodef{FFT}{fast Fourier transform}
\acrodef{FGS}{fine guidance sensor}
\acrodef{FOM}[FoM]{figure of merit}
\acrodef{FOM1}{folding mirror 1}
\acrodef{FOM2}{folding mirror 2}
\acrodef{FOM3}{folding mirror 3}
\acrodef{FOV}[FoV]{field of view}
\acrodef{FPA}{focal-plane array}
\acrodef{FWA}{filter-wheel assembly}
\acrodef{FWHM}{full width at half maximum}
\acrodef{GC}{galaxy clustering}
\acrodef{GFW}{grism and filter wheel}
\acrodef{H2RG}{HAWAII-2RG}
\acrodef{IMA}{image plane}
\acrodef{IQ}{image quality}
\acrodef{LAM}{Laboratoire d'Astrophysique de Marseille}
\acrodef{M1}{primary mirror}
\acrodef{M2}{secondary mirror}
\acrodef{M3}{tertiary mirror}
\acrodef{MLI}{multi-layer insulation}
\acrodef{MMU}{Mass Memory Unit}
\acrodef{MPS}{micro-propulsion system}
\acrodef{NASTRAN} {NASA Structural ANalysis}
\acrodef{JPL}{NASA Jet Propulsion Laboratory}
\acrodef{MaREA} {Multidisciplinary REentry Analysis} 
\acrodef{NIR}{near-infrared}
\acrodef{NISP}{Near-Infrared Spectrometer and Photometer}
\acrodef{PDC}{phase-diversity calibration}
\acrodef{PDR}{Preliminary Design Review}
\acrodef{PLM}{payload module}
\acrodef{PV}{performance-verification}
\acrodef{PSF}{point spread function}
\acrodef{QE}{quantum efficiency}
\acrodef{QR}{Qualification Review}
\acrodef{ROE}{readout electronics}
\acrodef{ROI}[RoI]{region of interest}
\acrodef{ROS}{reference observing sequence}
\acrodef{RSU}{readout shutter unit}
\acrodef{SAA}{Solar aspect angle}
\acrodef{SED}{spectral energy distribution}
\acrodef{SiC}{Silicon Carbide}
\acrodef{SNR}[S/N]{signal-to-noise ratio}
\acrodef{STM}{structural thermal model}
\acrodef{STOP}{structural-thermal-optical performance}
\acrodef{SVM}{service module}
\acrodef{TEMASE}{temperature mapping sensitivity}
\acrodef{TMA}{three-mirror anastigmat}
\acrodef{TMM}{thermal mathematical model}
\acrodef{VIS}{visible imager}
\acrodef{WFE}{wavefront error}
\acrodef{WL}{weak gravitational lensing}

\title{\Euclid}
\subtitle{Structural-Thermal-Optical Performance}    

\maxdeadcycles=500 
\author{Euclid Collaboration: A.~Anselmi\orcid{0000-0001-9897-9779}\inst{\ref{aff1}}
\and R.~Laureijs\thanks{\email{rene.laureijs@gmail.com}}\inst{\ref{aff2},\ref{aff3}}
\and G.~D.~Racca\orcid{0000-0002-9883-8981}\inst{\ref{aff3},\ref{aff4}}
\and G.~Costa\inst{\ref{aff5}}
\and L.~Courcould~Mifsud\inst{\ref{aff6}}
\and J.-C.~Cuillandre\orcid{0000-0002-3263-8645}\inst{\ref{aff7}}
\and M.~Gottero\inst{\ref{aff1}}
\and H.~Hoekstra\orcid{0000-0002-0641-3231}\inst{\ref{aff4}}
\and K.~Kuijken\orcid{0000-0002-3827-0175}\inst{\ref{aff4}}
\and V.~Mareschi\inst{\ref{aff1}}
\and L.~Miller\orcid{0000-0002-3376-6200}\inst{\ref{aff8}}
\and S.~Mottini\orcid{0000-0001-8327-2180}\inst{\ref{aff1}}
\and D.~Stramaccioni\inst{\ref{aff3}}
\and B.~Altieri\orcid{0000-0003-3936-0284}\inst{\ref{aff9}}
\and A.~Amara\inst{\ref{aff10}}
\and S.~Andreon\orcid{0000-0002-2041-8784}\inst{\ref{aff11}}
\and N.~Auricchio\orcid{0000-0003-4444-8651}\inst{\ref{aff12}}
\and C.~Baccigalupi\orcid{0000-0002-8211-1630}\inst{\ref{aff13},\ref{aff14},\ref{aff15},\ref{aff16}}
\and M.~Baldi\orcid{0000-0003-4145-1943}\inst{\ref{aff17},\ref{aff12},\ref{aff18}}
\and A.~Balestra\orcid{0000-0002-6967-261X}\inst{\ref{aff19}}
\and S.~Bardelli\orcid{0000-0002-8900-0298}\inst{\ref{aff12}}
\and R.~Bender\orcid{0000-0001-7179-0626}\inst{\ref{aff20},\ref{aff21}}
\and A.~Biviano\orcid{0000-0002-0857-0732}\inst{\ref{aff14},\ref{aff13}}
\and E.~Branchini\orcid{0000-0002-0808-6908}\inst{\ref{aff22},\ref{aff23},\ref{aff11}}
\and M.~Brescia\orcid{0000-0001-9506-5680}\inst{\ref{aff24},\ref{aff25}}
\and S.~Camera\orcid{0000-0003-3399-3574}\inst{\ref{aff26},\ref{aff27},\ref{aff28}}
\and G.~Ca\~nas-Herrera\orcid{0000-0003-2796-2149}\inst{\ref{aff3},\ref{aff4}}
\and V.~Capobianco\orcid{0000-0002-3309-7692}\inst{\ref{aff28}}
\and C.~Carbone\orcid{0000-0003-0125-3563}\inst{\ref{aff29}}
\and J.~Carretero\orcid{0000-0002-3130-0204}\inst{\ref{aff30},\ref{aff31}}
\and M.~Castellano\orcid{0000-0001-9875-8263}\inst{\ref{aff32}}
\and G.~Castignani\orcid{0000-0001-6831-0687}\inst{\ref{aff12}}
\and S.~Cavuoti\orcid{0000-0002-3787-4196}\inst{\ref{aff25},\ref{aff33}}
\and A.~Cimatti\inst{\ref{aff34}}
\and C.~Colodro-Conde\inst{\ref{aff35}}
\and G.~Congedo\orcid{0000-0003-2508-0046}\inst{\ref{aff36}}
\and C.~J.~Conselice\orcid{0000-0003-1949-7638}\inst{\ref{aff37}}
\and L.~Conversi\orcid{0000-0002-6710-8476}\inst{\ref{aff38},\ref{aff9}}
\and Y.~Copin\orcid{0000-0002-5317-7518}\inst{\ref{aff39}}
\and F.~Courbin\orcid{0000-0003-0758-6510}\inst{\ref{aff40},\ref{aff41}}
\and H.~M.~Courtois\orcid{0000-0003-0509-1776}\inst{\ref{aff42}}
\and M.~Cropper\orcid{0000-0003-4571-9468}\inst{\ref{aff43}}
\and A.~Da~Silva\orcid{0000-0002-6385-1609}\inst{\ref{aff44},\ref{aff45}}
\and H.~Degaudenzi\orcid{0000-0002-5887-6799}\inst{\ref{aff46}}
\and G.~De~Lucia\orcid{0000-0002-6220-9104}\inst{\ref{aff14}}
\and H.~Dole\orcid{0000-0002-9767-3839}\inst{\ref{aff47}}
\and F.~Dubath\orcid{0000-0002-6533-2810}\inst{\ref{aff46}}
\and F.~Ducret\inst{\ref{aff48}}
\and C.~A.~J.~Duncan\orcid{0009-0003-3573-0791}\inst{\ref{aff36}}
\and X.~Dupac\inst{\ref{aff9}}
\and S.~Dusini\orcid{0000-0002-1128-0664}\inst{\ref{aff49}}
\and S.~Escoffier\orcid{0000-0002-2847-7498}\inst{\ref{aff50}}
\and M.~Fabricius\orcid{0000-0002-7025-6058}\inst{\ref{aff20},\ref{aff21}}
\and M.~Farina\orcid{0000-0002-3089-7846}\inst{\ref{aff51}}
\and R.~Farinelli\inst{\ref{aff12}}
\and F.~Faustini\orcid{0000-0001-6274-5145}\inst{\ref{aff32},\ref{aff52}}
\and S.~Ferriol\inst{\ref{aff39}}
\and F.~Finelli\orcid{0000-0002-6694-3269}\inst{\ref{aff12},\ref{aff53}}
\and N.~Fourmanoit\orcid{0009-0005-6816-6925}\inst{\ref{aff50}}
\and M.~Frailis\orcid{0000-0002-7400-2135}\inst{\ref{aff14}}
\and E.~Franceschi\orcid{0000-0002-0585-6591}\inst{\ref{aff12}}
\and M.~Fumana\orcid{0000-0001-6787-5950}\inst{\ref{aff29}}
\and S.~Galeotta\orcid{0000-0002-3748-5115}\inst{\ref{aff14}}
\and K.~George\orcid{0000-0002-1734-8455}\inst{\ref{aff54}}
\and B.~Gillis\orcid{0000-0002-4478-1270}\inst{\ref{aff36}}
\and C.~Giocoli\orcid{0000-0002-9590-7961}\inst{\ref{aff12},\ref{aff18}}
\and J.~Gracia-Carpio\inst{\ref{aff20}}
\and A.~Grazian\orcid{0000-0002-5688-0663}\inst{\ref{aff19}}
\and F.~Grupp\inst{\ref{aff20},\ref{aff21}}
\and S.~V.~H.~Haugan\orcid{0000-0001-9648-7260}\inst{\ref{aff55}}
\and J.~Hoar\inst{\ref{aff9}}
\and W.~Holmes\inst{\ref{aff56}}
\and F.~Hormuth\inst{\ref{aff57}}
\and A.~Hornstrup\orcid{0000-0002-3363-0936}\inst{\ref{aff58},\ref{aff59}}
\and K.~Jahnke\orcid{0000-0003-3804-2137}\inst{\ref{aff60}}
\and M.~Jhabvala\inst{\ref{aff61}}
\and E.~Keih\"anen\orcid{0000-0003-1804-7715}\inst{\ref{aff62}}
\and S.~Kermiche\orcid{0000-0002-0302-5735}\inst{\ref{aff50}}
\and A.~Kiessling\orcid{0000-0002-2590-1273}\inst{\ref{aff56}}
\and R.~Kohley\inst{\ref{aff9}}
\and B.~Kubik\orcid{0009-0006-5823-4880}\inst{\ref{aff39}}
\and M.~Kunz\orcid{0000-0002-3052-7394}\inst{\ref{aff63}}
\and H.~Kurki-Suonio\orcid{0000-0002-4618-3063}\inst{\ref{aff64},\ref{aff65}}
\and A.~M.~C.~Le~Brun\orcid{0000-0002-0936-4594}\inst{\ref{aff66}}
\and S.~Ligori\orcid{0000-0003-4172-4606}\inst{\ref{aff28}}
\and P.~B.~Lilje\orcid{0000-0003-4324-7794}\inst{\ref{aff55}}
\and V.~Lindholm\orcid{0000-0003-2317-5471}\inst{\ref{aff64},\ref{aff65}}
\and I.~Lloro\orcid{0000-0001-5966-1434}\inst{\ref{aff67}}
\and G.~Mainetti\orcid{0000-0003-2384-2377}\inst{\ref{aff68}}
\and D.~Maino\inst{\ref{aff69},\ref{aff29},\ref{aff70}}
\and E.~Maiorano\orcid{0000-0003-2593-4355}\inst{\ref{aff12}}
\and O.~Mansutti\orcid{0000-0001-5758-4658}\inst{\ref{aff14}}
\and O.~Marggraf\orcid{0000-0001-7242-3852}\inst{\ref{aff71}}
\and M.~Martinelli\orcid{0000-0002-6943-7732}\inst{\ref{aff32},\ref{aff72}}
\and N.~Martinet\orcid{0000-0003-2786-7790}\inst{\ref{aff48}}
\and F.~Marulli\orcid{0000-0002-8850-0303}\inst{\ref{aff73},\ref{aff12},\ref{aff18}}
\and R.~J.~Massey\orcid{0000-0002-6085-3780}\inst{\ref{aff74}}
\and E.~Medinaceli\orcid{0000-0002-4040-7783}\inst{\ref{aff12}}
\and S.~Mei\orcid{0000-0002-2849-559X}\inst{\ref{aff75},\ref{aff76}}
\and Y.~Mellier\inst{\ref{aff77},\ref{aff78}}
\and M.~Meneghetti\orcid{0000-0003-1225-7084}\inst{\ref{aff12},\ref{aff18}}
\and E.~Merlin\orcid{0000-0001-6870-8900}\inst{\ref{aff32}}
\and G.~Meylan\inst{\ref{aff79}}
\and A.~Mora\orcid{0000-0002-1922-8529}\inst{\ref{aff80}}
\and M.~Moresco\orcid{0000-0002-7616-7136}\inst{\ref{aff73},\ref{aff12}}
\and L.~Moscardini\orcid{0000-0002-3473-6716}\inst{\ref{aff73},\ref{aff12},\ref{aff18}}
\and R.~Nakajima\orcid{0009-0009-1213-7040}\inst{\ref{aff71}}
\and C.~Neissner\orcid{0000-0001-8524-4968}\inst{\ref{aff81},\ref{aff31}}
\and R.~C.~Nichol\orcid{0000-0003-0939-6518}\inst{\ref{aff10}}
\and S.-M.~Niemi\orcid{0009-0005-0247-0086}\inst{\ref{aff3}}
\and C.~Padilla\orcid{0000-0001-7951-0166}\inst{\ref{aff81}}
\and S.~Paltani\orcid{0000-0002-8108-9179}\inst{\ref{aff46}}
\and F.~Pasian\orcid{0000-0002-4869-3227}\inst{\ref{aff14}}
\and K.~Pedersen\inst{\ref{aff82}}
\and W.~J.~Percival\orcid{0000-0002-0644-5727}\inst{\ref{aff83},\ref{aff84},\ref{aff85}}
\and V.~Pettorino\orcid{0000-0002-4203-9320}\inst{\ref{aff3}}
\and S.~Pires\orcid{0000-0002-0249-2104}\inst{\ref{aff7}}
\and G.~Polenta\orcid{0000-0003-4067-9196}\inst{\ref{aff52}}
\and M.~Poncet\inst{\ref{aff86}}
\and L.~A.~Popa\inst{\ref{aff87}}
\and F.~Raison\orcid{0000-0002-7819-6918}\inst{\ref{aff20}}
\and R.~Rebolo\orcid{0000-0003-3767-7085}\inst{\ref{aff35},\ref{aff88},\ref{aff89}}
\and A.~Renzi\orcid{0000-0001-9856-1970}\inst{\ref{aff90},\ref{aff49}}
\and J.~Rhodes\orcid{0000-0002-4485-8549}\inst{\ref{aff56}}
\and G.~Riccio\inst{\ref{aff25}}
\and E.~Romelli\orcid{0000-0003-3069-9222}\inst{\ref{aff14}}
\and M.~Roncarelli\orcid{0000-0001-9587-7822}\inst{\ref{aff12}}
\and C.~Rosset\orcid{0000-0003-0286-2192}\inst{\ref{aff75}}
\and E.~Rossetti\orcid{0000-0003-0238-4047}\inst{\ref{aff17}}
\and R.~Saglia\orcid{0000-0003-0378-7032}\inst{\ref{aff21},\ref{aff20}}
\and Z.~Sakr\orcid{0000-0002-4823-3757}\inst{\ref{aff91},\ref{aff92},\ref{aff93}}
\and J.-C.~Salvignol\inst{\ref{aff3}}
\and A.~G.~S\'anchez\orcid{0000-0003-1198-831X}\inst{\ref{aff20}}
\and D.~Sapone\orcid{0000-0001-7089-4503}\inst{\ref{aff94}}
\and B.~Sartoris\orcid{0000-0003-1337-5269}\inst{\ref{aff21},\ref{aff14}}
\and M.~Schirmer\orcid{0000-0003-2568-9994}\inst{\ref{aff60}}
\and P.~Schneider\orcid{0000-0001-8561-2679}\inst{\ref{aff71}}
\and T.~Schrabback\orcid{0000-0002-6987-7834}\inst{\ref{aff95}}
\and A.~Secroun\orcid{0000-0003-0505-3710}\inst{\ref{aff50}}
\and G.~Seidel\orcid{0000-0003-2907-353X}\inst{\ref{aff60}}
\and S.~Serrano\orcid{0000-0002-0211-2861}\inst{\ref{aff96},\ref{aff97},\ref{aff98}}
\and C.~Sirignano\orcid{0000-0002-0995-7146}\inst{\ref{aff90},\ref{aff49}}
\and G.~Sirri\orcid{0000-0003-2626-2853}\inst{\ref{aff18}}
\and J.~Skottfelt\orcid{0000-0003-1310-8283}\inst{\ref{aff99}}
\and L.~Stanco\orcid{0000-0002-9706-5104}\inst{\ref{aff49}}
\and J.~Steinwagner\orcid{0000-0001-7443-1047}\inst{\ref{aff20}}
\and P.~Tallada-Cresp\'{i}\orcid{0000-0002-1336-8328}\inst{\ref{aff30},\ref{aff31}}
\and D.~Tavagnacco\orcid{0000-0001-7475-9894}\inst{\ref{aff14}}
\and A.~N.~Taylor\inst{\ref{aff36}}
\and H.~I.~Teplitz\orcid{0000-0002-7064-5424}\inst{\ref{aff100}}
\and I.~Tereno\orcid{0000-0002-4537-6218}\inst{\ref{aff44},\ref{aff101}}
\and N.~Tessore\orcid{0000-0002-9696-7931}\inst{\ref{aff102}}
\and S.~Toft\orcid{0000-0003-3631-7176}\inst{\ref{aff103},\ref{aff104}}
\and R.~Toledo-Moreo\orcid{0000-0002-2997-4859}\inst{\ref{aff105}}
\and F.~Torradeflot\orcid{0000-0003-1160-1517}\inst{\ref{aff31},\ref{aff30}}
\and I.~Tutusaus\orcid{0000-0002-3199-0399}\inst{\ref{aff98},\ref{aff96},\ref{aff92}}
\and E.~A.~Valentijn\inst{\ref{aff2}}
\and L.~Valenziano\orcid{0000-0002-1170-0104}\inst{\ref{aff12},\ref{aff53}}
\and J.~Valiviita\orcid{0000-0001-6225-3693}\inst{\ref{aff64},\ref{aff65}}
\and T.~Vassallo\orcid{0000-0001-6512-6358}\inst{\ref{aff14},\ref{aff54}}
\and G.~Verdoes~Kleijn\orcid{0000-0001-5803-2580}\inst{\ref{aff2}}
\and A.~Veropalumbo\orcid{0000-0003-2387-1194}\inst{\ref{aff11},\ref{aff23},\ref{aff22}}
\and Y.~Wang\orcid{0000-0002-4749-2984}\inst{\ref{aff100}}
\and J.~Weller\orcid{0000-0002-8282-2010}\inst{\ref{aff21},\ref{aff20}}
\and A.~Zacchei\orcid{0000-0003-0396-1192}\inst{\ref{aff14},\ref{aff13}}
\and G.~Zamorani\orcid{0000-0002-2318-301X}\inst{\ref{aff12}}
\and E.~Zucca\orcid{0000-0002-5845-8132}\inst{\ref{aff12}}
\and M.~Ballardini\orcid{0000-0003-4481-3559}\inst{\ref{aff106},\ref{aff107},\ref{aff12}}
\and M.~Bolzonella\orcid{0000-0003-3278-4607}\inst{\ref{aff12}}
\and E.~Bozzo\orcid{0000-0002-8201-1525}\inst{\ref{aff46}}
\and C.~Burigana\orcid{0000-0002-3005-5796}\inst{\ref{aff108},\ref{aff53}}
\and R.~Cabanac\orcid{0000-0001-6679-2600}\inst{\ref{aff92}}
\and A.~Cappi\inst{\ref{aff12},\ref{aff109}}
\and J.~A.~Escartin~Vigo\inst{\ref{aff20}}
\and L.~Gabarra\orcid{0000-0002-8486-8856}\inst{\ref{aff8}}
\and W.~G.~Hartley\inst{\ref{aff46}}
\and J.~Mart\'{i}n-Fleitas\orcid{0000-0002-8594-569X}\inst{\ref{aff110}}
\and S.~Matthew\orcid{0000-0001-8448-1697}\inst{\ref{aff36}}
\and N.~Mauri\orcid{0000-0001-8196-1548}\inst{\ref{aff34},\ref{aff18}}
\and R.~B.~Metcalf\orcid{0000-0003-3167-2574}\inst{\ref{aff73},\ref{aff12}}
\and A.~Pezzotta\orcid{0000-0003-0726-2268}\inst{\ref{aff11}}
\and M.~P\"ontinen\orcid{0000-0001-5442-2530}\inst{\ref{aff64}}
\and I.~Risso\orcid{0000-0003-2525-7761}\inst{\ref{aff11},\ref{aff23}}
\and V.~Scottez\orcid{0009-0008-3864-940X}\inst{\ref{aff77},\ref{aff111}}
\and M.~Sereno\orcid{0000-0003-0302-0325}\inst{\ref{aff12},\ref{aff18}}
\and M.~Tenti\orcid{0000-0002-4254-5901}\inst{\ref{aff18}}
\and M.~Viel\orcid{0000-0002-2642-5707}\inst{\ref{aff13},\ref{aff14},\ref{aff16},\ref{aff15},\ref{aff112}}
\and M.~Wiesmann\orcid{0009-0000-8199-5860}\inst{\ref{aff55}}
\and Y.~Akrami\orcid{0000-0002-2407-7956}\inst{\ref{aff113},\ref{aff114}}
\and I.~T.~Andika\orcid{0000-0001-6102-9526}\inst{\ref{aff115},\ref{aff116}}
\and S.~Anselmi\orcid{0000-0002-3579-9583}\inst{\ref{aff49},\ref{aff90},\ref{aff117}}
\and M.~Archidiacono\orcid{0000-0003-4952-9012}\inst{\ref{aff69},\ref{aff70}}
\and F.~Atrio-Barandela\orcid{0000-0002-2130-2513}\inst{\ref{aff118}}
\and D.~Bertacca\orcid{0000-0002-2490-7139}\inst{\ref{aff90},\ref{aff19},\ref{aff49}}
\and M.~Bethermin\orcid{0000-0002-3915-2015}\inst{\ref{aff119}}
\and A.~Blanchard\orcid{0000-0001-8555-9003}\inst{\ref{aff92}}
\and L.~Blot\orcid{0000-0002-9622-7167}\inst{\ref{aff120},\ref{aff66}}
\and M.~Bonici\orcid{0000-0002-8430-126X}\inst{\ref{aff83},\ref{aff29}}
\and S.~Borgani\orcid{0000-0001-6151-6439}\inst{\ref{aff121},\ref{aff13},\ref{aff14},\ref{aff15},\ref{aff112}}
\and M.~L.~Brown\orcid{0000-0002-0370-8077}\inst{\ref{aff37}}
\and S.~Bruton\orcid{0000-0002-6503-5218}\inst{\ref{aff122}}
\and A.~Calabro\orcid{0000-0003-2536-1614}\inst{\ref{aff32}}
\and B.~Camacho~Quevedo\orcid{0000-0002-8789-4232}\inst{\ref{aff13},\ref{aff16},\ref{aff14}}
\and F.~Caro\inst{\ref{aff32}}
\and C.~S.~Carvalho\inst{\ref{aff101}}
\and T.~Castro\orcid{0000-0002-6292-3228}\inst{\ref{aff14},\ref{aff15},\ref{aff13},\ref{aff112}}
\and F.~Cogato\orcid{0000-0003-4632-6113}\inst{\ref{aff73},\ref{aff12}}
\and S.~Conseil\orcid{0000-0002-3657-4191}\inst{\ref{aff39}}
\and A.~R.~Cooray\orcid{0000-0002-3892-0190}\inst{\ref{aff123}}
\and O.~Cucciati\orcid{0000-0002-9336-7551}\inst{\ref{aff12}}
\and S.~Davini\orcid{0000-0003-3269-1718}\inst{\ref{aff23}}
\and G.~Desprez\orcid{0000-0001-8325-1742}\inst{\ref{aff2}}
\and A.~D\'iaz-S\'anchez\orcid{0000-0003-0748-4768}\inst{\ref{aff124}}
\and J.~J.~Diaz\orcid{0000-0003-2101-1078}\inst{\ref{aff35}}
\and S.~Di~Domizio\orcid{0000-0003-2863-5895}\inst{\ref{aff22},\ref{aff23}}
\and J.~M.~Diego\orcid{0000-0001-9065-3926}\inst{\ref{aff125}}
\and M.~Y.~Elkhashab\orcid{0000-0001-9306-2603}\inst{\ref{aff14},\ref{aff15},\ref{aff121},\ref{aff13}}
\and A.~Enia\orcid{0000-0002-0200-2857}\inst{\ref{aff12},\ref{aff17}}
\and Y.~Fang\orcid{0000-0002-0334-6950}\inst{\ref{aff21}}
\and A.~G.~Ferrari\orcid{0009-0005-5266-4110}\inst{\ref{aff18}}
\and A.~Finoguenov\orcid{0000-0002-4606-5403}\inst{\ref{aff64}}
\and A.~Franco\orcid{0000-0002-4761-366X}\inst{\ref{aff126},\ref{aff127},\ref{aff128}}
\and K.~Ganga\orcid{0000-0001-8159-8208}\inst{\ref{aff75}}
\and J.~Garc\'ia-Bellido\orcid{0000-0002-9370-8360}\inst{\ref{aff113}}
\and T.~Gasparetto\orcid{0000-0002-7913-4866}\inst{\ref{aff32}}
\and E.~Gaztanaga\orcid{0000-0001-9632-0815}\inst{\ref{aff98},\ref{aff96},\ref{aff129}}
\and F.~Giacomini\orcid{0000-0002-3129-2814}\inst{\ref{aff18}}
\and F.~Gianotti\orcid{0000-0003-4666-119X}\inst{\ref{aff12}}
\and G.~Gozaliasl\orcid{0000-0002-0236-919X}\inst{\ref{aff130},\ref{aff64}}
\and M.~Guidi\orcid{0000-0001-9408-1101}\inst{\ref{aff17},\ref{aff12}}
\and C.~M.~Gutierrez\orcid{0000-0001-7854-783X}\inst{\ref{aff131}}
\and A.~Hall\orcid{0000-0002-3139-8651}\inst{\ref{aff36}}
\and H.~Hildebrandt\orcid{0000-0002-9814-3338}\inst{\ref{aff132}}
\and J.~Hjorth\orcid{0000-0002-4571-2306}\inst{\ref{aff82}}
\and J.~J.~E.~Kajava\orcid{0000-0002-3010-8333}\inst{\ref{aff133},\ref{aff134}}
\and Y.~Kang\orcid{0009-0000-8588-7250}\inst{\ref{aff46}}
\and V.~Kansal\orcid{0000-0002-4008-6078}\inst{\ref{aff135},\ref{aff136}}
\and D.~Karagiannis\orcid{0000-0002-4927-0816}\inst{\ref{aff106},\ref{aff137}}
\and K.~Kiiveri\inst{\ref{aff62}}
\and J.~Kim\orcid{0000-0003-2776-2761}\inst{\ref{aff8}}
\and C.~C.~Kirkpatrick\inst{\ref{aff62}}
\and S.~Kruk\orcid{0000-0001-8010-8879}\inst{\ref{aff9}}
\and J.~Le~Graet\orcid{0000-0001-6523-7971}\inst{\ref{aff50}}
\and L.~Legrand\orcid{0000-0003-0610-5252}\inst{\ref{aff138},\ref{aff139}}
\and M.~Lembo\orcid{0000-0002-5271-5070}\inst{\ref{aff78}}
\and F.~Lepori\orcid{0009-0000-5061-7138}\inst{\ref{aff140}}
\and G.~Leroy\orcid{0009-0004-2523-4425}\inst{\ref{aff141},\ref{aff74}}
\and G.~F.~Lesci\orcid{0000-0002-4607-2830}\inst{\ref{aff73},\ref{aff12}}
\and J.~Lesgourgues\orcid{0000-0001-7627-353X}\inst{\ref{aff142}}
\and L.~Leuzzi\orcid{0009-0006-4479-7017}\inst{\ref{aff12}}
\and T.~I.~Liaudat\orcid{0000-0002-9104-314X}\inst{\ref{aff143}}
\and S.~J.~Liu\orcid{0000-0001-7680-2139}\inst{\ref{aff51}}
\and A.~Loureiro\orcid{0000-0002-4371-0876}\inst{\ref{aff144},\ref{aff145}}
\and J.~Macias-Perez\orcid{0000-0002-5385-2763}\inst{\ref{aff146}}
\and M.~Magliocchetti\orcid{0000-0001-9158-4838}\inst{\ref{aff51}}
\and F.~Mannucci\orcid{0000-0002-4803-2381}\inst{\ref{aff147}}
\and R.~Maoli\orcid{0000-0002-6065-3025}\inst{\ref{aff148},\ref{aff32}}
\and C.~J.~A.~P.~Martins\orcid{0000-0002-4886-9261}\inst{\ref{aff149},\ref{aff150}}
\and L.~Maurin\orcid{0000-0002-8406-0857}\inst{\ref{aff47}}
\and M.~Miluzio\inst{\ref{aff9},\ref{aff151}}
\and P.~Monaco\orcid{0000-0003-2083-7564}\inst{\ref{aff121},\ref{aff14},\ref{aff15},\ref{aff13},\ref{aff112}}
\and A.~Montoro\orcid{0000-0003-4730-8590}\inst{\ref{aff98},\ref{aff96}}
\and C.~Moretti\orcid{0000-0003-3314-8936}\inst{\ref{aff14},\ref{aff13},\ref{aff15},\ref{aff16}}
\and G.~Morgante\inst{\ref{aff12}}
\and S.~Nadathur\orcid{0000-0001-9070-3102}\inst{\ref{aff129}}
\and K.~Naidoo\orcid{0000-0002-9182-1802}\inst{\ref{aff129},\ref{aff102}}
\and A.~Navarro-Alsina\orcid{0000-0002-3173-2592}\inst{\ref{aff71}}
\and S.~Nesseris\orcid{0000-0002-0567-0324}\inst{\ref{aff113}}
\and D.~Paoletti\orcid{0000-0003-4761-6147}\inst{\ref{aff12},\ref{aff53}}
\and F.~Passalacqua\orcid{0000-0002-8606-4093}\inst{\ref{aff90},\ref{aff49}}
\and K.~Paterson\orcid{0000-0001-8340-3486}\inst{\ref{aff60}}
\and L.~Patrizii\inst{\ref{aff18}}
\and A.~Pisani\orcid{0000-0002-6146-4437}\inst{\ref{aff50}}
\and D.~Potter\orcid{0000-0002-0757-5195}\inst{\ref{aff140}}
\and S.~Quai\orcid{0000-0002-0449-8163}\inst{\ref{aff73},\ref{aff12}}
\and M.~Radovich\orcid{0000-0002-3585-866X}\inst{\ref{aff19}}
\and S.~Sacquegna\orcid{0000-0002-8433-6630}\inst{\ref{aff152},\ref{aff127},\ref{aff126}}
\and M.~Sahl\'en\orcid{0000-0003-0973-4804}\inst{\ref{aff153}}
\and D.~B.~Sanders\orcid{0000-0002-1233-9998}\inst{\ref{aff154}}
\and E.~Sarpa\orcid{0000-0002-1256-655X}\inst{\ref{aff16},\ref{aff112},\ref{aff15}}
\and A.~Schneider\orcid{0000-0001-7055-8104}\inst{\ref{aff140}}
\and D.~Sciotti\orcid{0009-0008-4519-2620}\inst{\ref{aff32},\ref{aff72}}
\and E.~Sellentin\inst{\ref{aff155},\ref{aff4}}
\and L.~C.~Smith\orcid{0000-0002-3259-2771}\inst{\ref{aff156}}
\and K.~Tanidis\orcid{0000-0001-9843-5130}\inst{\ref{aff8}}
\and G.~Testera\inst{\ref{aff23}}
\and R.~Teyssier\orcid{0000-0001-7689-0933}\inst{\ref{aff157}}
\and S.~Tosi\orcid{0000-0002-7275-9193}\inst{\ref{aff22},\ref{aff23},\ref{aff11}}
\and A.~Troja\orcid{0000-0003-0239-4595}\inst{\ref{aff90},\ref{aff49}}
\and M.~Tucci\inst{\ref{aff46}}
\and C.~Valieri\inst{\ref{aff18}}
\and A.~Venhola\orcid{0000-0001-6071-4564}\inst{\ref{aff158}}
\and D.~Vergani\orcid{0000-0003-0898-2216}\inst{\ref{aff12}}
\and G.~Verza\orcid{0000-0002-1886-8348}\inst{\ref{aff159}}
\and P.~Vielzeuf\orcid{0000-0003-2035-9339}\inst{\ref{aff50}}
\and N.~A.~Walton\orcid{0000-0003-3983-8778}\inst{\ref{aff156}}}
										   
\institute{Thales Alenia Space -- Euclid satellite Prime contractor, Strada Antica di Collegno 253, 10146 Torino, Italy\label{aff1}
\and
Kapteyn Astronomical Institute, University of Groningen, PO Box 800, 9700 AV Groningen, The Netherlands\label{aff2}
\and
European Space Agency/ESTEC, Keplerlaan 1, 2201 AZ Noordwijk, The Netherlands\label{aff3}
\and
Leiden Observatory, Leiden University, Einsteinweg 55, 2333 CC Leiden, The Netherlands\label{aff4}
\and
Aerospace Logistics Technology Engineering Company, Corso Marche 79, 10146 Torino, Italy\label{aff5}
\and
Airbus Defence \& Space SAS, Toulouse, France\label{aff6}
\and
Universit\'e Paris-Saclay, Universit\'e Paris Cit\'e, CEA, CNRS, AIM, 91191, Gif-sur-Yvette, France\label{aff7}
\and
Department of Physics, Oxford University, Keble Road, Oxford OX1 3RH, UK\label{aff8}
\and
ESAC/ESA, Camino Bajo del Castillo, s/n., Urb. Villafranca del Castillo, 28692 Villanueva de la Ca\~nada, Madrid, Spain\label{aff9}
\and
School of Mathematics and Physics, University of Surrey, Guildford, Surrey, GU2 7XH, UK\label{aff10}
\and
INAF-Osservatorio Astronomico di Brera, Via Brera 28, 20122 Milano, Italy\label{aff11}
\and
INAF-Osservatorio di Astrofisica e Scienza dello Spazio di Bologna, Via Piero Gobetti 93/3, 40129 Bologna, Italy\label{aff12}
\and
IFPU, Institute for Fundamental Physics of the Universe, via Beirut 2, 34151 Trieste, Italy\label{aff13}
\and
INAF-Osservatorio Astronomico di Trieste, Via G. B. Tiepolo 11, 34143 Trieste, Italy\label{aff14}
\and
INFN, Sezione di Trieste, Via Valerio 2, 34127 Trieste TS, Italy\label{aff15}
\and
SISSA, International School for Advanced Studies, Via Bonomea 265, 34136 Trieste TS, Italy\label{aff16}
\and
Dipartimento di Fisica e Astronomia, Universit\`a di Bologna, Via Gobetti 93/2, 40129 Bologna, Italy\label{aff17}
\and
INFN-Sezione di Bologna, Viale Berti Pichat 6/2, 40127 Bologna, Italy\label{aff18}
\and
INAF-Osservatorio Astronomico di Padova, Via dell'Osservatorio 5, 35122 Padova, Italy\label{aff19}
\and
Max Planck Institute for Extraterrestrial Physics, Giessenbachstr. 1, 85748 Garching, Germany\label{aff20}
\and
Universit\"ats-Sternwarte M\"unchen, Fakult\"at f\"ur Physik, Ludwig-Maximilians-Universit\"at M\"unchen, Scheinerstrasse 1, 81679 M\"unchen, Germany\label{aff21}
\and
Dipartimento di Fisica, Universit\`a di Genova, Via Dodecaneso 33, 16146, Genova, Italy\label{aff22}
\and
INFN-Sezione di Genova, Via Dodecaneso 33, 16146, Genova, Italy\label{aff23}
\and
Department of Physics "E. Pancini", University Federico II, Via Cinthia 6, 80126, Napoli, Italy\label{aff24}
\and
INAF-Osservatorio Astronomico di Capodimonte, Via Moiariello 16, 80131 Napoli, Italy\label{aff25}
\and
Dipartimento di Fisica, Universit\`a degli Studi di Torino, Via P. Giuria 1, 10125 Torino, Italy\label{aff26}
\and
INFN-Sezione di Torino, Via P. Giuria 1, 10125 Torino, Italy\label{aff27}
\and
INAF-Osservatorio Astrofisico di Torino, Via Osservatorio 20, 10025 Pino Torinese (TO), Italy\label{aff28}
\and
INAF-IASF Milano, Via Alfonso Corti 12, 20133 Milano, Italy\label{aff29}
\and
Centro de Investigaciones Energ\'eticas, Medioambientales y Tecnol\'ogicas (CIEMAT), Avenida Complutense 40, 28040 Madrid, Spain\label{aff30}
\and
Port d'Informaci\'{o} Cient\'{i}fica, Campus UAB, C. Albareda s/n, 08193 Bellaterra (Barcelona), Spain\label{aff31}
\and
INAF-Osservatorio Astronomico di Roma, Via Frascati 33, 00078 Monteporzio Catone, Italy\label{aff32}
\and
INFN section of Naples, Via Cinthia 6, 80126, Napoli, Italy\label{aff33}
\and
Dipartimento di Fisica e Astronomia "Augusto Righi" - Alma Mater Studiorum Universit\`a di Bologna, Viale Berti Pichat 6/2, 40127 Bologna, Italy\label{aff34}
\and
Instituto de Astrof\'{\i}sica de Canarias, V\'{\i}a L\'actea, 38205 La Laguna, Tenerife, Spain\label{aff35}
\and
Institute for Astronomy, University of Edinburgh, Royal Observatory, Blackford Hill, Edinburgh EH9 3HJ, UK\label{aff36}
\and
Jodrell Bank Centre for Astrophysics, Department of Physics and Astronomy, University of Manchester, Oxford Road, Manchester M13 9PL, UK\label{aff37}
\and
European Space Agency/ESRIN, Largo Galileo Galilei 1, 00044 Frascati, Roma, Italy\label{aff38}
\and
Universit\'e Claude Bernard Lyon 1, CNRS/IN2P3, IP2I Lyon, UMR 5822, Villeurbanne, F-69100, France\label{aff39}
\and
Institut de Ci\`{e}ncies del Cosmos (ICCUB), Universitat de Barcelona (IEEC-UB), Mart\'{i} i Franqu\`{e}s 1, 08028 Barcelona, Spain\label{aff40}
\and
Instituci\'o Catalana de Recerca i Estudis Avan\c{c}ats (ICREA), Passeig de Llu\'{\i}s Companys 23, 08010 Barcelona, Spain\label{aff41}
\and
UCB Lyon 1, CNRS/IN2P3, IUF, IP2I Lyon, 4 rue Enrico Fermi, 69622 Villeurbanne, France\label{aff42}
\and
Mullard Space Science Laboratory, University College London, Holmbury St Mary, Dorking, Surrey RH5 6NT, UK\label{aff43}
\and
Departamento de F\'isica, Faculdade de Ci\^encias, Universidade de Lisboa, Edif\'icio C8, Campo Grande, PT1749-016 Lisboa, Portugal\label{aff44}
\and
Instituto de Astrof\'isica e Ci\^encias do Espa\c{c}o, Faculdade de Ci\^encias, Universidade de Lisboa, Campo Grande, 1749-016 Lisboa, Portugal\label{aff45}
\and
Department of Astronomy, University of Geneva, ch. d'Ecogia 16, 1290 Versoix, Switzerland\label{aff46}
\and
Universit\'e Paris-Saclay, CNRS, Institut d'astrophysique spatiale, 91405, Orsay, France\label{aff47}
\and
Aix-Marseille Universit\'e, CNRS, CNES, LAM, Marseille, France\label{aff48}
\and
INFN-Padova, Via Marzolo 8, 35131 Padova, Italy\label{aff49}
\and
Aix-Marseille Universit\'e, CNRS/IN2P3, CPPM, Marseille, France\label{aff50}
\and
INAF-Istituto di Astrofisica e Planetologia Spaziali, via del Fosso del Cavaliere, 100, 00100 Roma, Italy\label{aff51}
\and
Space Science Data Center, Italian Space Agency, via del Politecnico snc, 00133 Roma, Italy\label{aff52}
\and
INFN-Bologna, Via Irnerio 46, 40126 Bologna, Italy\label{aff53}
\and
University Observatory, LMU Faculty of Physics, Scheinerstrasse 1, 81679 Munich, Germany\label{aff54}
\and
Institute of Theoretical Astrophysics, University of Oslo, P.O. Box 1029 Blindern, 0315 Oslo, Norway\label{aff55}
\and
Jet Propulsion Laboratory, California Institute of Technology, 4800 Oak Grove Drive, Pasadena, CA, 91109, USA\label{aff56}
\and
Felix Hormuth Engineering, Goethestr. 17, 69181 Leimen, Germany\label{aff57}
\and
Technical University of Denmark, Elektrovej 327, 2800 Kgs. Lyngby, Denmark\label{aff58}
\and
Cosmic Dawn Center (DAWN), Denmark\label{aff59}
\and
Max-Planck-Institut f\"ur Astronomie, K\"onigstuhl 17, 69117 Heidelberg, Germany\label{aff60}
\and
NASA Goddard Space Flight Center, Greenbelt, MD 20771, USA\label{aff61}
\and
Department of Physics and Helsinki Institute of Physics, Gustaf H\"allstr\"omin katu 2, University of Helsinki, 00014 Helsinki, Finland\label{aff62}
\and
Universit\'e de Gen\`eve, D\'epartement de Physique Th\'eorique and Centre for Astroparticle Physics, 24 quai Ernest-Ansermet, CH-1211 Gen\`eve 4, Switzerland\label{aff63}
\and
Department of Physics, P.O. Box 64, University of Helsinki, 00014 Helsinki, Finland\label{aff64}
\and
Helsinki Institute of Physics, Gustaf H{\"a}llstr{\"o}min katu 2, University of Helsinki, 00014 Helsinki, Finland\label{aff65}
\and
Laboratoire d'etude de l'Univers et des phenomenes eXtremes, Observatoire de Paris, Universit\'e PSL, Sorbonne Universit\'e, CNRS, 92190 Meudon, France\label{aff66}
\and
SKAO, Jodrell Bank, Lower Withington, Macclesfield SK11 9FT, UK\label{aff67}
\and
Centre de Calcul de l'IN2P3/CNRS, 21 avenue Pierre de Coubertin 69627 Villeurbanne Cedex, France\label{aff68}
\and
Dipartimento di Fisica "Aldo Pontremoli", Universit\`a degli Studi di Milano, Via Celoria 16, 20133 Milano, Italy\label{aff69}
\and
INFN-Sezione di Milano, Via Celoria 16, 20133 Milano, Italy\label{aff70}
\and
Universit\"at Bonn, Argelander-Institut f\"ur Astronomie, Auf dem H\"ugel 71, 53121 Bonn, Germany\label{aff71}
\and
INFN-Sezione di Roma, Piazzale Aldo Moro, 2 - c/o Dipartimento di Fisica, Edificio G. Marconi, 00185 Roma, Italy\label{aff72}
\and
Dipartimento di Fisica e Astronomia "Augusto Righi" - Alma Mater Studiorum Universit\`a di Bologna, via Piero Gobetti 93/2, 40129 Bologna, Italy\label{aff73}
\and
Department of Physics, Institute for Computational Cosmology, Durham University, South Road, Durham, DH1 3LE, UK\label{aff74}
\and
Universit\'e Paris Cit\'e, CNRS, Astroparticule et Cosmologie, 75013 Paris, France\label{aff75}
\and
CNRS-UCB International Research Laboratory, Centre Pierre Bin\'etruy, IRL2007, CPB-IN2P3, Berkeley, USA\label{aff76}
\and
Institut d'Astrophysique de Paris, 98bis Boulevard Arago, 75014, Paris, France\label{aff77}
\and
Institut d'Astrophysique de Paris, UMR 7095, CNRS, and Sorbonne Universit\'e, 98 bis boulevard Arago, 75014 Paris, France\label{aff78}
\and
Institute of Physics, Laboratory of Astrophysics, Ecole Polytechnique F\'ed\'erale de Lausanne (EPFL), Observatoire de Sauverny, 1290 Versoix, Switzerland\label{aff79}
\and
Telespazio UK S.L. for European Space Agency (ESA), Camino bajo del Castillo, s/n, Urbanizacion Villafranca del Castillo, Villanueva de la Ca\~nada, 28692 Madrid, Spain\label{aff80}
\and
Institut de F\'{i}sica d'Altes Energies (IFAE), The Barcelona Institute of Science and Technology, Campus UAB, 08193 Bellaterra (Barcelona), Spain\label{aff81}
\and
DARK, Niels Bohr Institute, University of Copenhagen, Jagtvej 155, 2200 Copenhagen, Denmark\label{aff82}
\and
Waterloo Centre for Astrophysics, University of Waterloo, Waterloo, Ontario N2L 3G1, Canada\label{aff83}
\and
Department of Physics and Astronomy, University of Waterloo, Waterloo, Ontario N2L 3G1, Canada\label{aff84}
\and
Perimeter Institute for Theoretical Physics, Waterloo, Ontario N2L 2Y5, Canada\label{aff85}
\and
Centre National d'Etudes Spatiales -- Centre spatial de Toulouse, 18 avenue Edouard Belin, 31401 Toulouse Cedex 9, France\label{aff86}
\and
Institute of Space Science, Str. Atomistilor, nr. 409 M\u{a}gurele, Ilfov, 077125, Romania\label{aff87}
\and
Consejo Superior de Investigaciones Cientificas, Calle Serrano 117, 28006 Madrid, Spain\label{aff88}
\and
Universidad de La Laguna, Departamento de Astrof\'{\i}sica, 38206 La Laguna, Tenerife, Spain\label{aff89}
\and
Dipartimento di Fisica e Astronomia "G. Galilei", Universit\`a di Padova, Via Marzolo 8, 35131 Padova, Italy\label{aff90}
\and
Institut f\"ur Theoretische Physik, University of Heidelberg, Philosophenweg 16, 69120 Heidelberg, Germany\label{aff91}
\and
Institut de Recherche en Astrophysique et Plan\'etologie (IRAP), Universit\'e de Toulouse, CNRS, UPS, CNES, 14 Av. Edouard Belin, 31400 Toulouse, France\label{aff92}
\and
Universit\'e St Joseph; Faculty of Sciences, Beirut, Lebanon\label{aff93}
\and
Departamento de F\'isica, FCFM, Universidad de Chile, Blanco Encalada 2008, Santiago, Chile\label{aff94}
\and
Universit\"at Innsbruck, Institut f\"ur Astro- und Teilchenphysik, Technikerstr. 25/8, 6020 Innsbruck, Austria\label{aff95}
\and
Institut d'Estudis Espacials de Catalunya (IEEC),  Edifici RDIT, Campus UPC, 08860 Castelldefels, Barcelona, Spain\label{aff96}
\and
Satlantis, University Science Park, Sede Bld 48940, Leioa-Bilbao, Spain\label{aff97}
\and
Institute of Space Sciences (ICE, CSIC), Campus UAB, Carrer de Can Magrans, s/n, 08193 Barcelona, Spain\label{aff98}
\and
Centre for Electronic Imaging, Open University, Walton Hall, Milton Keynes, MK7~6AA, UK\label{aff99}
\and
Infrared Processing and Analysis Center, California Institute of Technology, Pasadena, CA 91125, USA\label{aff100}
\and
Instituto de Astrof\'isica e Ci\^encias do Espa\c{c}o, Faculdade de Ci\^encias, Universidade de Lisboa, Tapada da Ajuda, 1349-018 Lisboa, Portugal\label{aff101}
\and
Department of Physics and Astronomy, University College London, Gower Street, London WC1E 6BT, UK\label{aff102}
\and
Cosmic Dawn Center (DAWN)\label{aff103}
\and
Niels Bohr Institute, University of Copenhagen, Jagtvej 128, 2200 Copenhagen, Denmark\label{aff104}
\and
Universidad Polit\'ecnica de Cartagena, Departamento de Electr\'onica y Tecnolog\'ia de Computadoras,  Plaza del Hospital 1, 30202 Cartagena, Spain\label{aff105}
\and
Dipartimento di Fisica e Scienze della Terra, Universit\`a degli Studi di Ferrara, Via Giuseppe Saragat 1, 44122 Ferrara, Italy\label{aff106}
\and
Istituto Nazionale di Fisica Nucleare, Sezione di Ferrara, Via Giuseppe Saragat 1, 44122 Ferrara, Italy\label{aff107}
\and
INAF, Istituto di Radioastronomia, Via Piero Gobetti 101, 40129 Bologna, Italy\label{aff108}
\and
Universit\'e C\^{o}te d'Azur, Observatoire de la C\^{o}te d'Azur, CNRS, Laboratoire Lagrange, Bd de l'Observatoire, CS 34229, 06304 Nice cedex 4, France\label{aff109}
\and
Aurora Technology for European Space Agency (ESA), Camino bajo del Castillo, s/n, Urbanizacion Villafranca del Castillo, Villanueva de la Ca\~nada, 28692 Madrid, Spain\label{aff110}
\and
ICL, Junia, Universit\'e Catholique de Lille, LITL, 59000 Lille, France\label{aff111}
\and
ICSC - Centro Nazionale di Ricerca in High Performance Computing, Big Data e Quantum Computing, Via Magnanelli 2, Bologna, Italy\label{aff112}
\and
Instituto de F\'isica Te\'orica UAM-CSIC, Campus de Cantoblanco, 28049 Madrid, Spain\label{aff113}
\and
CERCA/ISO, Department of Physics, Case Western Reserve University, 10900 Euclid Avenue, Cleveland, OH 44106, USA\label{aff114}
\and
Technical University of Munich, TUM School of Natural Sciences, Physics Department, James-Franck-Str.~1, 85748 Garching, Germany\label{aff115}
\and
Max-Planck-Institut f\"ur Astrophysik, Karl-Schwarzschild-Str.~1, 85748 Garching, Germany\label{aff116}
\and
Laboratoire Univers et Th\'eorie, Observatoire de Paris, Universit\'e PSL, Universit\'e Paris Cit\'e, CNRS, 92190 Meudon, France\label{aff117}
\and
Departamento de F{\'\i}sica Fundamental. Universidad de Salamanca. Plaza de la Merced s/n. 37008 Salamanca, Spain\label{aff118}
\and
Universit\'e de Strasbourg, CNRS, Observatoire astronomique de Strasbourg, UMR 7550, 67000 Strasbourg, France\label{aff119}
\and
Center for Data-Driven Discovery, Kavli IPMU (WPI), UTIAS, The University of Tokyo, Kashiwa, Chiba 277-8583, Japan\label{aff120}
\and
Dipartimento di Fisica - Sezione di Astronomia, Universit\`a di Trieste, Via Tiepolo 11, 34131 Trieste, Italy\label{aff121}
\and
California Institute of Technology, 1200 E California Blvd, Pasadena, CA 91125, USA\label{aff122}
\and
Department of Physics \& Astronomy, University of California Irvine, Irvine CA 92697, USA\label{aff123}
\and
Departamento F\'isica Aplicada, Universidad Polit\'ecnica de Cartagena, Campus Muralla del Mar, 30202 Cartagena, Murcia, Spain\label{aff124}
\and
Instituto de F\'isica de Cantabria, Edificio Juan Jord\'a, Avenida de los Castros, 39005 Santander, Spain\label{aff125}
\and
INFN, Sezione di Lecce, Via per Arnesano, CP-193, 73100, Lecce, Italy\label{aff126}
\and
Department of Mathematics and Physics E. De Giorgi, University of Salento, Via per Arnesano, CP-I93, 73100, Lecce, Italy\label{aff127}
\and
INAF-Sezione di Lecce, c/o Dipartimento Matematica e Fisica, Via per Arnesano, 73100, Lecce, Italy\label{aff128}
\and
Institute of Cosmology and Gravitation, University of Portsmouth, Portsmouth PO1 3FX, UK\label{aff129}
\and
Department of Computer Science, Aalto University, PO Box 15400, Espoo, FI-00 076, Finland\label{aff130}
\and
Instituto de Astrof\'\i sica de Canarias, c/ Via Lactea s/n, La Laguna 38200, Spain. Departamento de Astrof\'\i sica de la Universidad de La Laguna, Avda. Francisco Sanchez, La Laguna, 38200, Spain\label{aff131}
\and
Ruhr University Bochum, Faculty of Physics and Astronomy, Astronomical Institute (AIRUB), German Centre for Cosmological Lensing (GCCL), 44780 Bochum, Germany\label{aff132}
\and
Department of Physics and Astronomy, Vesilinnantie 5, University of Turku, 20014 Turku, Finland\label{aff133}
\and
Serco for European Space Agency (ESA), Camino bajo del Castillo, s/n, Urbanizacion Villafranca del Castillo, Villanueva de la Ca\~nada, 28692 Madrid, Spain\label{aff134}
\and
ARC Centre of Excellence for Dark Matter Particle Physics, Melbourne, Australia\label{aff135}
\and
Centre for Astrophysics \& Supercomputing, Swinburne University of Technology,  Hawthorn, Victoria 3122, Australia\label{aff136}
\and
Department of Physics and Astronomy, University of the Western Cape, Bellville, Cape Town, 7535, South Africa\label{aff137}
\and
DAMTP, Centre for Mathematical Sciences, Wilberforce Road, Cambridge CB3 0WA, UK\label{aff138}
\and
Kavli Institute for Cosmology Cambridge, Madingley Road, Cambridge, CB3 0HA, UK\label{aff139}
\and
Department of Astrophysics, University of Zurich, Winterthurerstrasse 190, 8057 Zurich, Switzerland\label{aff140}
\and
Department of Physics, Centre for Extragalactic Astronomy, Durham University, South Road, Durham, DH1 3LE, UK\label{aff141}
\and
Institute for Theoretical Particle Physics and Cosmology (TTK), RWTH Aachen University, 52056 Aachen, Germany\label{aff142}
\and
IRFU, CEA, Universit\'e Paris-Saclay 91191 Gif-sur-Yvette Cedex, France\label{aff143}
\and
Oskar Klein Centre for Cosmoparticle Physics, Department of Physics, Stockholm University, Stockholm, SE-106 91, Sweden\label{aff144}
\and
Astrophysics Group, Blackett Laboratory, Imperial College London, London SW7 2AZ, UK\label{aff145}
\and
Univ. Grenoble Alpes, CNRS, Grenoble INP, LPSC-IN2P3, 53, Avenue des Martyrs, 38000, Grenoble, France\label{aff146}
\and
INAF-Osservatorio Astrofisico di Arcetri, Largo E. Fermi 5, 50125, Firenze, Italy\label{aff147}
\and
Dipartimento di Fisica, Sapienza Universit\`a di Roma, Piazzale Aldo Moro 2, 00185 Roma, Italy\label{aff148}
\and
Centro de Astrof\'{\i}sica da Universidade do Porto, Rua das Estrelas, 4150-762 Porto, Portugal\label{aff149}
\and
Instituto de Astrof\'isica e Ci\^encias do Espa\c{c}o, Universidade do Porto, CAUP, Rua das Estrelas, PT4150-762 Porto, Portugal\label{aff150}
\and
HE Space for European Space Agency (ESA), Camino bajo del Castillo, s/n, Urbanizacion Villafranca del Castillo, Villanueva de la Ca\~nada, 28692 Madrid, Spain\label{aff151}
\and
INAF - Osservatorio Astronomico d'Abruzzo, Via Maggini, 64100, Teramo, Italy\label{aff152}
\and
Theoretical astrophysics, Department of Physics and Astronomy, Uppsala University, Box 516, 751 37 Uppsala, Sweden\label{aff153}
\and
Institute for Astronomy, University of Hawaii, 2680 Woodlawn Drive, Honolulu, HI 96822, USA\label{aff154}
\and
Mathematical Institute, University of Leiden, Einsteinweg 55, 2333 CA Leiden, The Netherlands\label{aff155}
\and
Institute of Astronomy, University of Cambridge, Madingley Road, Cambridge CB3 0HA, UK\label{aff156}
\and
Department of Astrophysical Sciences, Peyton Hall, Princeton University, Princeton, NJ 08544, USA\label{aff157}
\and
Space physics and astronomy research unit, University of Oulu, Pentti Kaiteran katu 1, FI-90014 Oulu, Finland\label{aff158}
\and
Center for Computational Astrophysics, Flatiron Institute, 162 5th Avenue, 10010, New York, NY, USA\label{aff159}}    

\date{Version 2.0 - October 2025}

 
  \abstract
   {The \Euclid system performance is defined in terms of image quality metrics tuned to the weak gravitational lensing cosmological probe. The weak lensing measurement induces stringent requirements on the shape and stability of the VIS instrument system \ac{PSF}. The \ac{PSF} is affected by error contributions from the telescope, the focal plane and image motion, and is controlled by a global error budget with error allocations to each contributor.
   }
   {During spacecraft development, we verified through a structural-thermal-optical performance (STOP) analysis that the built and verified telescope with its spacecraft interface meets the in-orbit steady-state and transient image quality requirements, under temperature-induced loads, in all permitted spacecraft attitudes after all permitted attitude transitions. In the first year in orbit, we compared the expected with the actual performance.}
   {For the purposes of the STOP analysis, a detailed finite-element mathematical model was set up and a standard set of test cases, both steady-state and transient, was defined, comprising combinations of worst-case boundary conditions. Iterations of the analysis were performed in conjunction with the major reviews of the spacecraft verification cycle. After launch, the model is used in sensitivity analyses using realistic boundary conditions.}
   {The STOP analysis addressed the interaction of all spacecraft components in transmitting temperature-induced loads that lead to optical train deformation. The results of the prelaunch analysis demonstrated that temperature-induced optical perturbations will be well below the allowable limits for all permitted observing conditions. During the first year in orbit, we used the STOP analysis predictions to help interpret the measured performance as a function of environmental variables. Unpredicted disturbances were discovered (heat pulses from instrument operation propagating into the telescope) and unexpected sensitivities were revealed (high dependence of $T_{\rm BP}$ on \ac{SAA}, nearly absent dependence on \ac{AA} after the attitude domain was redefined for straylight avoidance). In-orbit temperature variations are small (<300\,mK) and so are their effects on the telescope structure (displacements < 1 $\mu$m, rotations < 1 $\mu$rad), but they are detected in the time histories of the image quality metrics and are a non-negligible factor in the \ac{PSF} stability budget demanded by the weak lensing science ($\Delta e < 2{\times}10^{-3}$ over 11\,000\,s). Taking everything into account, our analysis confirms the excellent overall performance of the telescope.}
   {}


%
%
%
   \maketitle
%

\section{\label{sec:intro}Introduction}

\Euclid is a space-based survey mission to improve our understanding of the nature of dark energy and dark matter with an unprecedented level of accuracy and precision \citep{EuclidSkyOverview}. The mission has been optimised for the measurement of two dark energy probes: \ac{WL} and \ac{GC}. These complementary probes involve a statistical analysis of galaxies detected in \Euclid's survey area of 14\,000\,deg$^2$, which covers a major fraction of the extra-galactic sky that can be observed in the visible and near-infrared. To meet the high-precision cosmology objectives, a top-down design approach was adopted, where the science requirements were translated into a complete set of technical, operations, calibration, and data processing requirements. To deal with the complexity of budgeting and performance allocation, model-based system engineering was applied, see \cite{Lorenzo2016} and \cite{Vavrek2016}.

The \ac{WL} probe is based on the determination of galaxy shear from the measurement of the shape of galaxies derived from their size and ellipticity. The shear determination demands tight control of systematic effects down to the depth and resolution achievable; see \cite{Amara2008}, \cite{Paulin-Henriksson2008}, \cite{Paulin-Henriksson2009}, and \cite{Cropper2013} for a detailed assessment of the \Euclid case. The required properties of the \ac{PSF} detectable with the \ac{VIS} resulted in a set of \ac{IQ} requirements that have a major impact on the design of \Euclid, imposing high stability of mechanical structures, temperatures, and telescope pointing \citep{Racca2016}. The design aims to provide the maximum knowledge of the \ac{PSF} needed for the correction of the sources of systematic effects, whose aggregated error contribution should be significantly less than the statistical errors in the total error budget \citep{Cropper2013}.

\begin{figure}
    \centering
    \includegraphics[width=0.55\columnwidth]{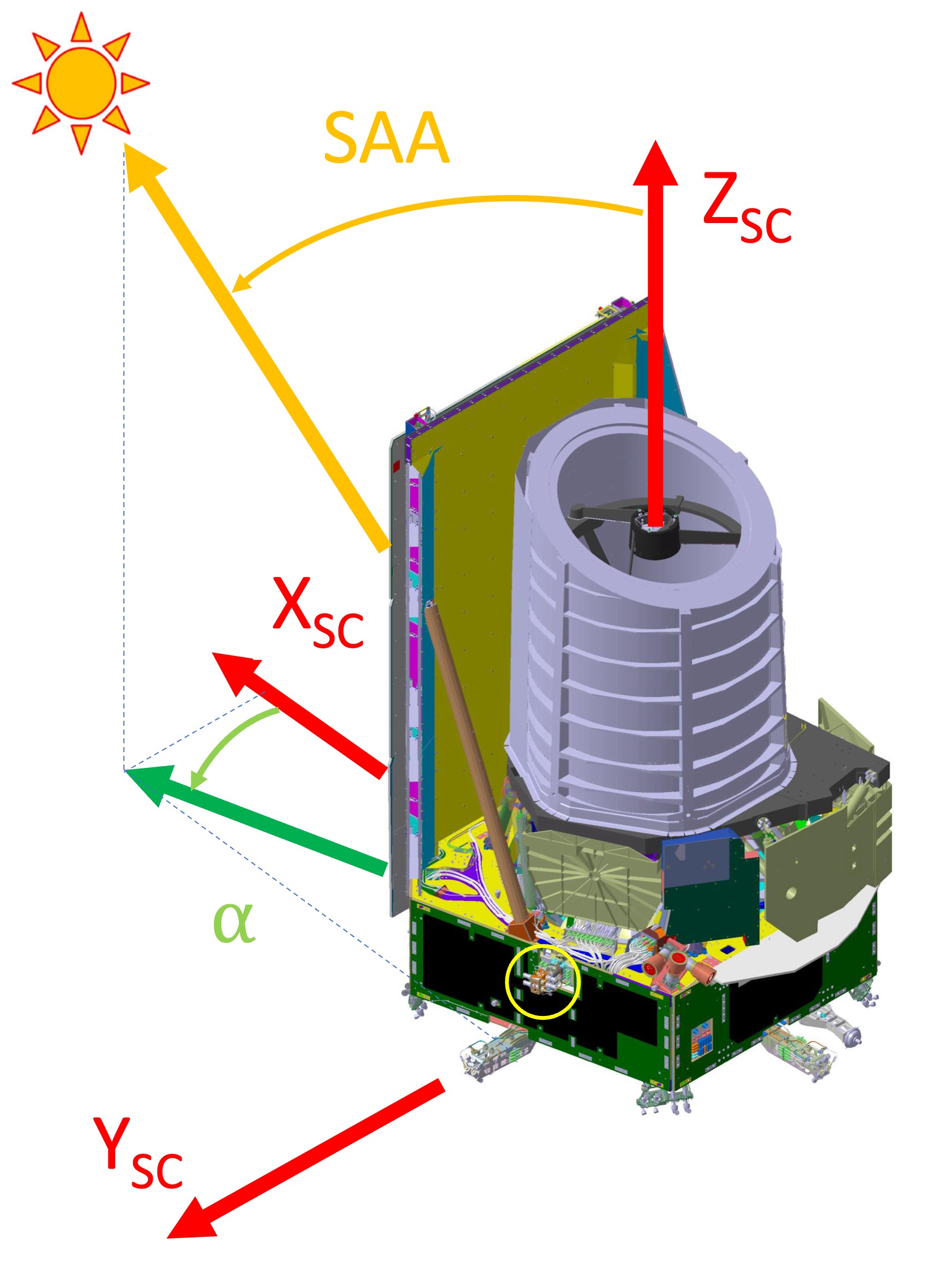}
    \caption{Definition of Solar aspect angle (SAA) and $\alpha$ angle (AA) with respect to the spacecraft axes. SAA is the angle between the Sun direction vector and the Z axis of the spacecraft, while AA is the angle between the X axis and the projection of the Sun direction on the X-Y plane. The spacecraft is a schematic presentation to illustrate the configuration without the protective multi-layer insulation. Thermal disturbances occur whenever SAA $> 90\degree$ where the Sun illuminates the bottom of the spacecraft, or AA $> -2{\fdg}5$ where the Sun illuminates the micro-propulsion system (MPS) boom, encircled in the figure.}
\label{FigAtt}
\end{figure}

Despite the very stable conditions that can be achieved in space, there can still be transient \ac{IQ} variations induced by short- and medium-term thermal variations. These are caused by changes in Solar attitude during the execution of the survey plan and variations in heat dissipation by electronic units. At the design level, countermeasures taken include high-efficiency thermal insulation, the use of materials with a low coefficient of thermal expansion, and carefully managed spacecraft operational modes to minimise variations in power dissipation. The designed Solar attitude range of the telescope is limited to angles $-$8\degree < AA < 8\degree \,and 87\degree < SAA < 121\degree \,(see \cref{FigAtt}).

As part of the design and development process, the adequacy of the \Euclid design was verified by dedicated component and instrument model analyses complemented by test campaigns after assembly and integration of subsystems, and eventually the entire system; see \cite{Gaspar2022}. To ensure the compliance of the \Euclid design with the \ac{IQ} requirements, we have developed the \ac{STOP} analysis, predicting the effects of thermal loads on the \ac{IQ} stability during \Euclid's operation in-orbit. An overview of the thermal and mechanical design and the development steps relevant for the analysis described in this document is provided in Appendix\,\ref{appendix:sec:design-summary}.

In this paper, we describe the image quality performance of \Euclid for the \ac{WL} probe as expected from the \ac{STOP} analysis, and we compare the \ac{STOP} predictions with the \ac{IQ} metrics measured during early \Euclid operations. \Cref{sec:requirements} discusses the requirements and the accompanying image quality metrics. \Cref{sec:approach} describes the approach to the analysis, the mathematical models, the model integration into the \ac{STOP} model suite and the model validation. \Cref{sec:design-phase} gives details of the thermal, thermoelastic, and optical analyses, and predictions about the satellite-level performance in terms of IQ metrics. \Cref{sec:commissioning} provides details of the early in-orbit performance and describes some ancillary analyses performed to support satellite commissioning. Finally, in \cref{sec:discussion} we compare the performance with the predictions and in \cref{sec:conclusions} we draw the conclusions and give suggestions for the application of STOP analysis during in-orbit operations.

\section{\label{sec:requirements}IQ performance requirements}

\begin{table}
    \caption{\Euclid image quality requirements for \ac{WL} measurements, the values give the maximum limits. The science requirements refer to the knowledge of the \ac{PSF} at a fiducial wavelength of 800 nm and a maximum integration time of 700\,s. The spacecraft requirements (column 3) give the derived allocations for the spacecraft. The science and spacecraft requirements for the bottom two lines refer to a period 700\,s and 11\,000\,s, respectively.}
    \centering          
    \begin{tabular}{l c c}     
        \hline\hline
        \noalign{\vskip 3pt}
        Variable & Science req. & Spacecraft req. \\
        \noalign{\smallskip}
        \hline 
        \noalign{\smallskip}
        FWHM [\si{\arcsecond}]           & 0.18  & 0.155 \\
        Ellipticity             & 0.15  & 0.14  \\ 
        $R^2$ [arcsec$^2$]      & 4$R_{\rm ref}^2$\,$^\dagger$  & 0.055 \\              
        $\delta e_i$       &   $2\times 10^{-4}$    & $2 \times 10^{-3}$ \\
        $\delta$$R^2$/$R^2$     & $10^{-3}$      & $2 \times 10^{-3}$ \\
        \noalign{\smallskip}
        \hline                  
    \end{tabular}
    \tablefoot{\\$\dagger$~$R_{\rm ref}^2$ is based on a Gaussian profile with a FWHM of $0\farcs2$. }
\label{tableReq}      
\end{table}

The spacecraft requirements applicable to the STOP analysis were derived from the science requirements for the \ac{WL} image quality, taking into account the apportionment to spacecraft, calibration, data processing, and survey strategy. The allocation to the spacecraft design was driven by the practical realisation of the design, by the expected performance of the intended in-flight calibrations, and the capability of the \ac{PSF} model. The spacecraft requirements include contributions from the design residuals, integration alignment error, focus error, thermal stability, and spacecraft-induced image motion.

The size of the \ac{PSF} is defined as the \ac{FWHM} averaged azimuthally about the \ac{PSF} centroid.
\begin{equation}
{\rm FWHM} = \frac{1}{\pi} \int_0^\pi {\rm FWHM}(\alpha)\;{\rm d} \alpha \;.
\end{equation}
The ellipticity $\it{\bf{\textit{e}}}$ is defined in terms of the Gaussian-weighted quadrupole moments, where $e = \sqrt{e_1^2 + e_2^2}$, and $e_1$ and $e_2$ are the two projected ellipticity components obtained from
\begin{equation}
e_1 = \frac{Q_{xx}-Q_{yy}}{Q_{xx}+Q_{yy}}\;, \quad e_2 = \frac{2Q_{xy}}{Q_{xx}+Q_{yy}}\;,
\end{equation}
where $Q_{ij}$ are the weighted quadrupole moments, as defined in \cite{Schneider2006} that involve a Gaussian weighting function with a standard deviation $\sigma =0{\farcs}75$, based on the typical size of the smallest galaxies useful for weak lensing \citep{Meneghetti2008}.
The \ac{IQ} metric $R^2=Q_{xx}+Q_{yy}$ bounds the \ac{PSF} wings to a reference value $R_{\rm ref}^2$ obtained from a Gaussian profile with \ac{FWHM} of $0{\farcs}2$. The spatio-temporal variation of the \ac{PSF} shall allow a model of the \ac{PSF} as a function of time and position in the image plane
with a residual model uncertainty smaller than $\delta e_i$ and $\delta R^2/R^2$.

The science requirements and the apportioned spacecraft requirements are listed in Table 1. Besides the static requirements for a fiducial exposure time of 700 s, it also provides the transient requirements for the spacecraft imposing a temporal stability for $\delta R^2/R^2$ and the ellipticity components. The temporal stability of the spacecraft is required to improve the model of the \ac{PSF} by aggregating stars from more than one field \citep{Cropper2013}. The spacecraft requirements are defined between any two of eight consecutive dither exposures of two adjacent survey fields \citep{Scaramella-EP1}, with a maximum duration of 11\,000\,s for 8 consecutive dither exposures. The corresponding science requirements refer to the knowledge over an exposure time of maximum 700\,s and therefore have smaller values. In the remainder of this paper, we will refer only to the spacecraft requirements.

\section{\label{sec:approach}STOP analysis approach}

The \ac{STOP} analysis considers all thermal loads that affect the stability of the telescope in-orbit. The thermal loads acting on the telescope come from its environment, which include the Sun, \ac{PLM} containing the telescope and scientific instruments, \ac{SVM} including electronic units, and the sunshield. The \ac{STOP} analysis shall predict the impact on the \ac{IQ} parameters, and the resulting distortion-driven error terms are considered in the system performance budget. The results of the analysis during the design phase were used to improve the system design as part of an iterative process toward the final design.

The design of the spacecraft must meet its requirements in all possible conditions during the lifetime of the mission. To define design cases, extreme conditions are considered for the permissible attitudes with respect to the Sun, for changes of thermo-optical properties of the exposed surfaces because the Solar absorptivity of the paint is expected to increase with age, and for electronic unit power dissipation taking ageing into account. The situations analysed are therefore conservative and relative to compounded worst cases.
 
Each iteration of the analysis proceeds through three coordinated sets of mathematical models according to the following sequence: temperature sets at appropriate “thermal” locations $\rightarrow$ interpolation to corresponding “structural” locations $\rightarrow$ displacement and distortion of the optical path and elements $\rightarrow$ application to the optical model $\rightarrow$ resulting optical aberrations (IQ metrics).

\subsection{Mathematical models}

The system \ac{TMM} was assembled from models representing each major component: \ac{SVM}, \ac{PLM}, and sunshield. The component models were prepared and tested separately. For the \ac{PLM}, two configurations were provided, operational worst-hot and operational worst-cold. In the system analysis, the worst-cold configuration was associated with beginning-of-life conditions and the worst-hot configuration with end-of-life conditions, due to the degradation of the thermo-optical properties of the surfaces.

The \ac{STOP} structural analysis \ac{FEM} considers the model assembly of the major components using MSC NASTRAN software (2008 and 2016 versions). The \ac{PLM} and sunshield are attached to the \ac{SVM} structure by means of rigid elements and bars. There are no structural links between \ac{PLM} and sunshield. The total mass of the model is 1974\,kg.
Customization of the \ac{FEM} for thermo-elastic analysis included the application of thermal expansion coefficients. The \ac{CTE} can vary with temperature and for some materials the difference between room temperature and the operating temperature of \Euclid can be large, e.g. the \ac{CTE} of \ac{SiC} used in the \ac{PLM} structures changes by a factor of about 4. For these sensitive materials, \ac{CTE} values measured at a nominal operating temperature of 135\,K were applied. 

\begin{figure}
    \centering
    \includegraphics[width=0.7\columnwidth]{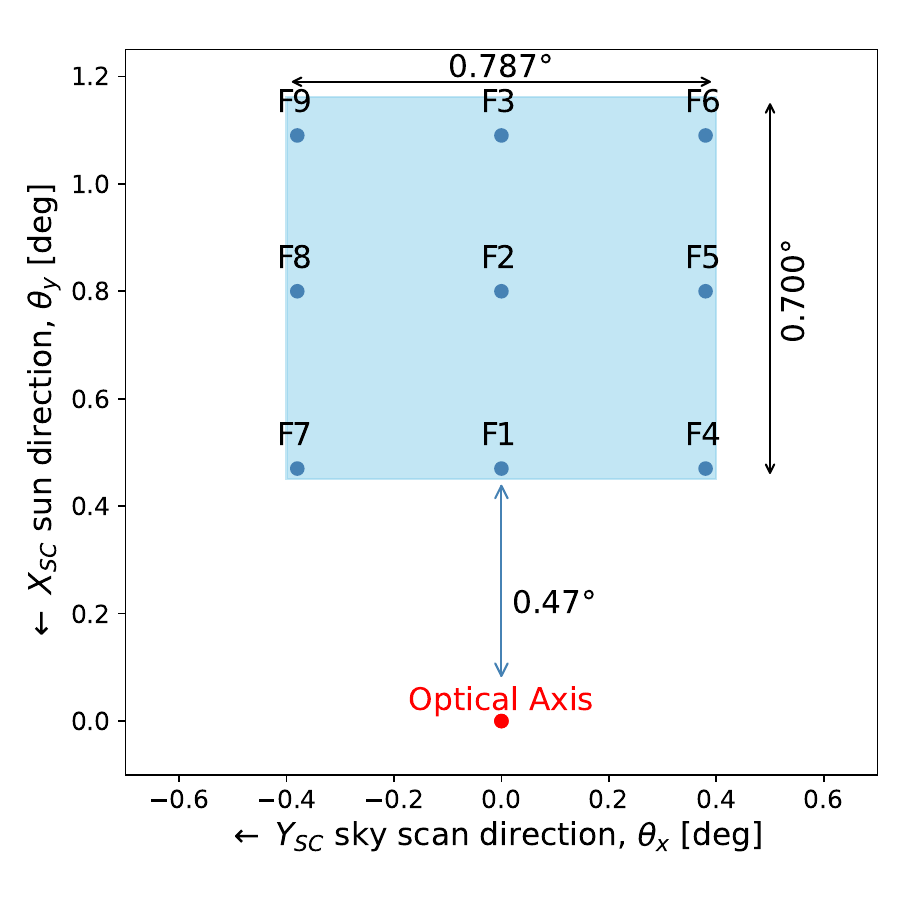}
    \caption{VIS image plane field points. The Korsch telescope design provides an off-axis exit pupil. See \cref{FigAtt} for the definition of the spacecraft $X$, $Y$, and $Z$ coordinates.}
\label{FigIMA}
\end{figure}

The optical model consists of the telescope aperture and the \ac{VIS} image plane, via the optical train composed of the \ac{M1}, \ac{M2}, \ac{FOM1}, \ac{FOM2}, \ac{M3}, dichroic filter, and \ac{FOM3}, see \cite{EuclidSkyOverview} for a schematic.  The \ac{PSF} was computed for 9 sample points in the \ac{VIS} focal plane by ray tracing, see Fig.\,\ref{FigIMA}. In the ray-tracing, the mirrors are represented by the location of their vertices, translated and rotated according to the thermo-elastic results. The effect of temperature on the curvature of the mirrors, whose main effect is defocus, was added using the analytical procedure outlined in Appendix \ref{appendix:sub:codeV}.
Ray tracing optical design software Code V\footnote{see https://www.synopsys.com/optical-solutions/codev.html} was customized to process the displacements and rotations provided by the structural analysis. All coordinates are referred to a global reference frame centred on the \ac{M1} vertex. Numerical noise in the computation of the ellipticity, the most sensitive figure, due to small differences in pupil sampling, is on the order of a few parts in \(10^4\). Other contributors, such as small differences between the actual coded algorithms, are one order of magnitude below.

The STOP analysis used the \ac{MaREA} software package, developed by Thales Alenia Space, to efficiently transfer data between mathematical models of different types. Using MaREA functions, the three mathematical models (thermal \ac{TMM}, structural \ac{FEM} and optical CODE V) were integrated into one workflow. The process was then implemented in the SIMULIA Isight® execution engine\footnote{SIMULIA Isight® is a commercial process automation and simulation tool developed by Dassault Systèmes.} for sequential runs of multiple analysis cases and statistical post-processing of the results. See Appendix \ref{appendix:sub:MaREA-description} for a detailed description.

\subsection{Definition of analysis cases}

\begin{table}
   \centering
    \caption[]{Definition of variables for steady-state analysis cases.}
    \begin{tabular}{ll}
        \hline\hline
        \noalign{\vskip 3pt}
        Variable                &  Value(s)  \\
        \noalign{\smallskip}
        \hline
        \noalign{\smallskip}
        Solar irradiance: & \\
		\,\,\,BOL, EOL$^{*}$   & 1293 Wm$^{-2}$, 1388 Wm$^{-2}$\\
        SAA  & 87{\degree}, 90{\degree}, 105{\degree}, 121{\degree}\\
        AA    & 8{\degree}, 0{\degree}, $-$8{\degree} \\
        Operational Mode        & NOM, COM$^{**}$ \\
        SVM top floor interfaces:          & \\
        \,\,\,conductive &  20\,{\degree}C, 24\,{\degree}C\\
        \,\,\,radiative &  worst-case MLI$^{***}$\\   
        Sunshield to PLM interface:          & \\
        \,\,\,radiative &  worst-case MLI\\   
        \noalign{\smallskip}
        \hline
    \end{tabular}
    \tablefoot{\\$*$ begin of life (BOL) and end of life (EOL) conditions \\$**$ Nominal (NOM) and Communications (COM) modes \\$***$ Multi Layer Insulation}
\label{TabTstat}
\end{table}

\begin{table}
   \centering
   \caption[]{STOP Transient sizing case (hottest to coldest).}
   \begin{tabular}{ll}
        \hline\hline
        \noalign{\vskip 3pt}
        Variable                &  Value  \\
        \noalign{\smallskip}
        \hline
        \noalign{\smallskip}
        Solar irradiance & 1388 Wm$^{-2}$ (EOL) \\
        PLM Operational Mode & OpHotWorst \\  
        Start SAA & 121\degree \\
        Start AA  & 8\degree\\
        Start SVM Operational Mode & COM \\
        End SAA & 87\degree \\
        End AA & $-$8\degree \\
        End SVM Operational Mode & NOM \\          
		\noalign{\smallskip}
		\hline
   \end{tabular}
\label{TabTtra}
\end{table}

IQ parameters were determined for a number of possible in-orbit situations in which the telescope is in a thermally stable state. The variables and their values for the steady-state cases are provided in \cref{TabTstat}. We obtained the IQ parameters for each thermal combination involving four Solar aspect angles (SAAs) and three $\alpha$ angles, for two operational modes: nominal (NOM) and higher dissipation communications (COM) mode, two values of the conductive interface, and two Solar irradiances, resulting in 96 cases. 

The transient case is intended for the verification of the image quality stability requirements (ellipticity and $R^2$). Table \ref{TabTtra} provides the definition of the transient case, covering the transition from the hottest to the coldest condition (EOL, at constant Solar irradiance), as allowed by the Solar illumination requirements. A mode transition of the SVM is included to envelope the extreme worst cases of environment variation, but is not representative of any real-life case, where the changes in attitudes and dissipation are always minimised.

\section{\label{sec:design-phase}STOP analysis in the satellite design phase}

\subsection{\label{sub:thermal-analysis}Thermal analysis}

The thermal response of the telescope structure to changes in the environment provides already a strong indication of the possible \ac{IQ} changes which are quantified in the subsequent \ac{STOP} analysis steps. A general property consistently observed is the linear correlation between the temperatures of the main optical elements and the average temperature of the telescope baseplate. This property of the baseplate is due to its large mass and surface area. The baseplate accounts for more than half of the mass of the SiC parts in the telescope dominating the telescope thermal capacity. In addition, the baseplate has the largest surface area facing the SVM, making it the main component for the radiative exchange between the PLM and the SVM. 

The PLM settings (dissipation, heater power) are important in determining the telescope temperatures and, to a lesser extent, their variations. All thermal states were calculated from an initial state with a uniform 20\,\degree C temperature in all thermal nodes. Consequently, the calculated deformations included the large cool-down component caused by the temperature transition from 20\,\degree C to the operating temperature of each thermal node. The common cool-down component was removed in the structural analysis by taking the difference of each state vector (translation and rotation) with a reference state vector associated with STOP case \#19
(SAA = 90\degree,  AA = 0\degree, interface temperature = 20\,\degree C , nominal mode, see also \cref{subsub:steady-state-thermal}).

\subsubsection{\label{subsub:steady-state-thermal}Steady-state thermal analysis}

\begin{figure}
    \centering
    \includegraphics[width=0.85\columnwidth]{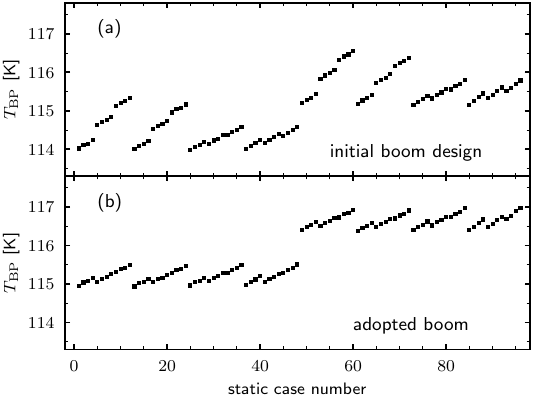}
    \caption{Baseplate average temperature versus STOP case number, see \cref{subsub:steady-state-thermal} for the sequencing description. {\bf Upper panel (a):} full-length \ac{MPS} boom, the topside can be illuminated by the Sun. {\bf Lower panel (b):} half-length \ac{MPS} boom, the topside cannot be illuminated by the Sun.}
\label{FigTBplot}
\end{figure}

The steady-state analysis addresses the 96 STOP cases defined in \cref{TabTstat}. The results are presented in plots of temperature versus STOP case number using the following convention.
\begin{itemize}
    \item The first set of 48 points refers to \ac{BOL} conditions and the second set of 48 points (\#49 to \#96) refers to \ac{EOL} conditions. \ac{BOL} is obtained from the minimum yearly irradiance at summer solstice at L2 distance from the Sun, and \ac{EOL} from the irradiance at winter solstice at L2.
    \item In both the \ac{BOL} set and the \ac{EOL} set, the first set of 12 points is calculated at \ac{SAA} = 87\degree, the second set at \ac{SAA}  = 90\degree, the third set at \ac{SAA}  = 105\degree \,and the fourth set at \ac{SAA}  = 121\degree.
    \item Within each set of 12 points defined above, the first four are characterised by \ac{AA} = $-$8\degree \,(Sun on the spacecraft \(-Y\)\ side), the second set of four by \ac{AA} = 0\degree, and the third set by \ac{AA} = +8\degree \,(Sun on \(+Y\)\  side).
    \item Within each subset of four consecutive points, the first two share the NOM operational mode and differ by the conductive interface temperature, 20\,\degree C or 24\,\degree C, and the second two share the COM operational mode and again differ by the conductive interface temperature, 20\,\degree C or 24\,\degree C.
\end{itemize}
From the resulting \cref{FigTBplot} one can trace the dependence on SAA (increasing left to right) and on azimuth (increasing bottom to top).

Figure \ref{FigTBplot}a shows the average temperature of the baseplate $T_\mathrm{BP}$ for the 96 steady-state thermal cases using the initial design configuration of \Euclid. For \ac{SAA} = 87\degree, an increase of $T_\mathrm{BP}$ as large as 1.4 K was found when rotating the spacecraft from AA=$-$8\degree \,to AA=+8\degree. Such a large temperature change was not expected. Initial estimates indicated that the ellipticity would change by 5\% per K temperature variation in $T_\mathrm{BP}$, the derived variations would violate the \ac{IQ} requirement for ellipticity.

The dependence of $T_\mathrm{BP}$ on Sun azimuth was attributed to the presence of the \ac{MPS} boom on the spacecraft  $+Y_\mathrm{SC}$ side, in front of and below the \ac{VIS} radiator. Radiation from Sun-facing side of the boom reflected off the back of the sunshield and reached the telescope baffle, which is mounted on the baseplate, see \cref{appendix:FigTmodplot} in Appendix\,\ref{appendix:sub:MPS-boom-redesign}. This finding led to a redesign of the \ac{MPS} boom. In a first approach, the \ac{MLI} on the boom topside was reconfigured such that the topside remained in the shadow. For the final configuration, the boom length was reduced to half its initial length (\cref{FigTBplot}b). This design solution would still give sufficient torque authority without causing a significant increase of $T_\mathrm{BP}$ compared to a solution without boom. The boom redesign is described in Appendix\,\ref{appendix:sub:MPS-boom-redesign}.
		
\subsubsection{\label{subsub:transient-thermal}Transient thermal analysis} 

\begin{figure}
    \centering
    \includegraphics[width=0.85\columnwidth]{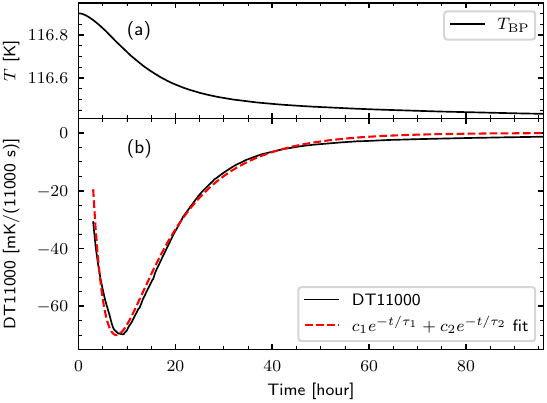}
    \caption{Temperature transient analysis. {\bf Upper panel (a):} baseplate temperature evolution over 96 hours. {\bf Lower panel (b):} the temperature difference DT11000 derived from the temperature evolution in (a); the dashed curve shows a match of DT11000 with two exponential terms with $c_1=$ 168 mK and $c_2=$ 315 mK.}
\label{FigTplotTra}
\end{figure}

We simulated a transient case that should provide maximum stress to the \ac{PLM} by imposing extreme changes; see Table \ref{TabTtra}. Assuming \ac{EOL}, we define an attitude transition simultaneously comprising the maximum allowed ranges of SAA and AA. The transition brings the Sun from the upper limit of the allowed \ac{SAA} (121\degree), a warm condition for the SVM, to the lowest \ac{SAA} (87\degree), a colder condition. The SVM warming is due to the illumination of the bottom platform by the Sun whenever \ac{SAA}\,>\,90\degree~ (cf. \cref{FigAtt}).
A mode change from COM to NOM is also included, causing a high to low variation in power dissipation in the \ac{SVM}. The transition causes a significant cooling of the \ac{SVM} and a slight warming of the \ac{PLM}. 

In the calculation, the attitude transition is instantaneous, applied at $t=0$, and the subsequent evolution is tracked for four days. It should be noted that four days are not enough to reach a steady condition, we estimate a drop below 0.1\% after seven days assuming exponential decay and a time constant of one day; therefore, the end state differs somewhat from that found in the corresponding case of the steady-state analysis. However, four days are long enough to capture all the essential features of the transient.

The temperature inputs in the \ac{IQ} analysis are the temperature differences over a 11\,000\,s sliding time window: DT11000 = $T(t)-T(t-11\,000\,{\rm s})$. \Cref{FigTplotTra}a shows the telescope baseplate temperature evolution over 96\,h after the attitude transition and  \cref{FigTplotTra}b shows the corresponding DT11000 for the case stipulated in Table \ref{TabTtra}. The dip in the DT11000 curve, lowest after about 8\,h, is explained as due to the interaction of two temperature components that act in a SAA slew. The displacement of the Sun from \ac{SAA} = 121\degree\, to 87\degree\, cools the \ac{SVM} and heats the sunshield. The two effects propagate to the PLM baseplate with different time constants. The superposition of the two effects produces the dip. The DT11000 curve is well matched by the combination of two exponential functions with time constants of $\tau_1$ = 12.3\,h, associated with the \ac{PLM}, and $\tau_2$ = 2.8\,h, associated with the \ac{SVM}, see \cref{FigTplotTra}b.

A sensitivity analysis was performed to study the dependence of the temperature perturbation on the environmental variables and the spacecraft properties. The results are summarised as follows. The smaller the attitude variation, the lower the temperature fluctuation and the minimum fluctuation is achieved when limiting SAA to less than 105\degree, and the azimuth range to $-$5\degree\,< AA < 5\degree. Minimising the change in power dissipation in the \ac{SVM}, too, reduces the magnitude of the temperature fluctuation. Reversing the start and end points of the transition described in Table \ref{TabTtra} does not produce a symmetric temperature change, and the magnitude of DT11000 is lower in the ``warming'' transition. The properties of the surface finishes have a small but noticeable effect on the position of the peak and its duration, which may be relevant in the case of ageing. 

\subsection{Structural analysis}

\begin{figure}
    \centering
    \includegraphics[width=0.85\columnwidth]{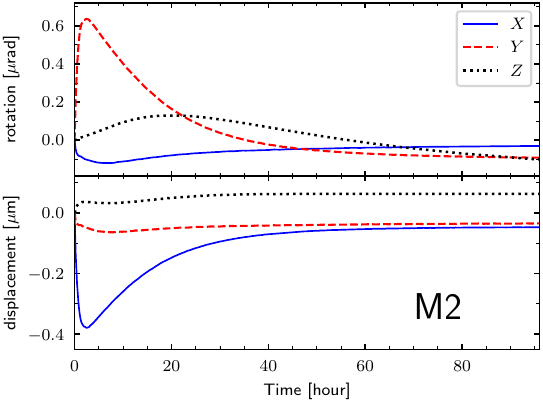}
    \caption{M2 rotations (upper panel) and displacements (lower panel) around the spacecraft axes after a transition according to the parameters in Table \ref{TabTtra}.} 
\label{FigDplotTra}
\end{figure}

The temperature variations from the thermal analysis are applied to the corresponding elements of the structure to obtain the displacements and rotations induced in the telescope optical train. The structural analysis addresses both the 96 steady-state temperature perturbations defined in \cref{subsub:steady-state-thermal} and the transient perturbations in \cref{subsub:transient-thermal}.

\subsubsection{\label{subsub:steady-state-structural}Steady-state structural analysis}

The elements of the optical train comprise 8 points representing the centre positions of the main mirrors \ac{M1}, \ac{M2}, and \ac{M3}, the folding mirrors \ac{FOM1}, \ac{FOM2}, and \ac{FOM3}, the dichroic, and the image plane.

The solid-body motion is removed by associating each element to the reference frame defined by \ac{M1}. Since the temperature perturbations are computed for each case from a uniform temperature of 20\,\degree C, they contain the ground-to-orbit temperature transition, which leads to a large deformation. This is removed in-orbit by focussing the telescope using the \ac{M2} mechanism. This operation is simulated by taking the difference of each state with that associated to a reference case, as described in \cref{sub:thermal-analysis}. Taking the differences is justified in a linear approximation, given the smallness of the effects once the common-mode motions have been removed. 

In all nodes, the maximum magnitude of each component of the displacement vector is found to be less than 1\,µm, and the maximum rotation of the reference vector is less than 1\,$\mu$rad. These displacements and rotations become significant in the optical analysis in \cref{sub:optical-analysis}.

\subsubsection{\label{subsub:stability-structural}Structural stability analysis }
We have computed the displacements and rotations of the optical elements with respect to their reference positions as a function of time for 4 days following the attitude transition defined in Table \ref{TabTtra}. All optical elements exhibit a rapid change in the first few hours after the transition, concurrent with the temperature perturbation of the baseplate described in \cref{subsub:transient-thermal}. The change appears as a damped oscillation, with a quick rise and then reversing direction with a slow exponential return. See, for example, the $y$-component of \ac{M2}, displayed in \cref{FigDplotTra}, with peak magnitude 0.65\,$\mu$rad.

\begin{table}
    \caption[]{Metrics of the “as built” telescope optical model.}
    \begin{tabular}{*{4}{c}}
        \hline\hline
        \noalign{\vskip 3pt}
        Field Point & Ellipticity & $R^2$ (arcsec²) & FWHM (\si{\arcsecond}) \\
        \hline
        \noalign{\smallskip}
        F1 & 0.005  & 0.046  & 0.1344 \\
        F2 & 0.007  & 0.045  & 0.1338 \\
        F3 & 0.014  & 0.047  & 0.1340 \\
        F4 & 0.027  & 0.051  & 0.1327 \\
        F5 & 0.033  & 0.046  & 0.1342 \\
        F6 & 0.020  & 0.051  & 0.1362 \\
        F7 & 0.044  & 0.049  & 0.1355 \\
        F8 & 0.050  & 0.047  & 0.1347 \\
        F9 & 0.025  & 0.049  & 0.1353 \\
        \noalign{\smallskip}
        \hline                  
    \end{tabular}
\label{TabOpTel}
\end{table}

\subsection{\label{sub:optical-analysis}Optical analysis}

For optical performance calculations of the \Euclid telescope, the \ac{PSF} at the desired points were generated by the CODE V `as built' optical model based on the nominal optical layout constructed after correlation with the results of the \ac{PLM} thermal vacuum test; see Appendix\,\ref{appendix:sec:design-summary}. The \ac{PSF} is then sampled with a pitch of $4\,\mu{\rm m}$ and processed using a $2048 \times 2048$ \ac{FFT}. The map is then truncated to $256 \times 256$ points centred on the maximum value. The truncation is justified by the fact that none of the weighted IQ metrics (ellipticity, $R^2$, and \ac{FWHM}) is significantly dependent on the part of the \ac{PSF} energy that is beyond $128 \times 4\,\mu{\rm m} = 512\,\mu{\rm m}$ or 4$\farcs$26 from its centre. The image quality metrics are calculated using a discrete version of the equations given in \cref{sec:requirements}. Table \ref{TabOpTel} shows the metrics of the nominal telescope realization. 
This process is time-consuming, and not suited for the STOP analysis where many different static cases, as well as temporal transient cases, have to be analysed. To ease and quicken the process, simplified models have been used, as described in Appendix\,\ref{appendix:sub:codeV}.

\subsubsection{\label{subsub:IQ-static}System steady-state IQ performance}

The \ac{IQ} metrics were calculated for the 96 cases defined in \cref{TabTstat}, both without and with the inclusion of the predicted deformation of the mirrors.We find that the influence of the telescope optical train deformation is very small and that the mirrors contribute to almost all the variance among the cases.

The different cases indicate notable variations in ellipticity due to changes in azimuth and operating mode. Little variation is caused by the different SAA. All variations are well within 0.5\% for the contribution by the optical train deformation alone. The deformation by the mirrors amplifies the ellipticity variations to about 1.0\% for the most perturbed field points F2, F6, F7, and F9 (cf. Fig. \ref{FigIMA}).
The most perturbed field points are well within the maximum allowed ellipticity (0.14). A similar behaviour is found for $R^2$. The \ac{FWHM} is hardly affected by thermo-elastic perturbations. We point out that $R^2$ is more sensitive to the PSF wings than the FWHM.

\subsubsection{\label{subsub:IQ-transient}System in-orbit IQ stability performance} 

\begin{figure}
    \centering
    \includegraphics[width=0.9\columnwidth]{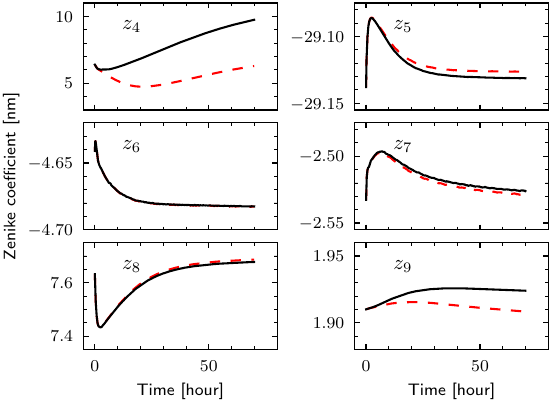}
    \caption{11\,000\,s  variation of Zernike coefficients $z_4$ to $z_9$. Dashed lines: telescope only; solid lines: telescope and mirrors.}
\label{FigOpZer}
\end{figure}

The stability of the telescope metrics is evaluated starting from the time series of displacements/rotations of the optical train described in \cref{subsub:stability-structural}, originating in the extreme thermal transient described in \cref{subsub:transient-thermal}. The transients were interrupted after 4 days, in a condition close to the final steady state ($\Delta T$ with respect to the corresponding static STOP case < 0.1~\degree C in all relevant nodes). The optical metrics were calculated up to 70\,h, for reasons of containing the time-consuming computations, but the essentials of the transients are clear anyway. See Appendix\,\ref{appendix:IQ-perf-calculations} for the computed timelines.

To understand the underlying variations in the shape of the \ac{PSF} causing the \ac{IQ} metrics, we analyse Zernike components $z_4$ to $z_9$ (adopting Fringe's ordering and notation) from which the \ac{PSF} can be constructed. The variations of the components are presented in \cref{FigOpZer}. It shows that the largest variations are found in the defocus term ($z_4$) followed by astigmatism ($z_5$) and coma ($z_8$), both of which are already an order of magnitude smaller than $z_4$. The additional defocus contribution due to mirror deformation (cf. Appendix\,\ref{appendix:sub:codeV}) increases monotonically towards an asymptotic value and dominates the value of $z_4$ near the end of the transient. The $z_4$ telescope term has a minimum after about 20 hours and then goes back to its original value. The mirror term correlates with the mean \ac{M1} temperature, whereas the telescope term correlates with the vertical displacement of \ac{M2} relative to \ac{M1}. Changes in the curvature of the mirrors, in particular M1, affect the focus, which is the dominant term. This justifies the telescope re-alignment using only M2 adjustments. Sensitivity analysis shows that virtually all ellipticity variation is cancelled when the \ac{M2} $\Delta Z$ variable is constrained to zero.

The higher order terms $z_5$ and $z_8$ both exhibit a fast transient with almost the same magnitude but opposite signs. The transient peaks after about 2\,h and then returns to its original value, with a small offset, after 24\,h. This transient can be modelled as the sum of two opposing terms with time constants of about 10\,h and about 1\,h (cf. \cref{subsub:transient-thermal}). The 10 h constant is associated with the baseplate. The fast component can be attributed to the telescope truss, as suggested by the strong correlation with the lateral \ac{M2} displacement $\Delta Y$. The short time constant can be explained by a strip of thermal nodes associated with the external baffle and VIS radiator with a view factor to the back of the sunshield (see Appendix\,\ref{appendix:sub:MPS-boom-redesign}) and changing its temperature on a timescale of less than one hour. This heat is dissipated via the baseplate.

\begin{table}
    \caption{\Euclid Image Quality performance (VIS channel). The STOP performance values were calculated before launch; they are worst cases and do not belong to the same realisation. The in-orbit ellipticity and FWHM values were derived from a VIS frame taken from the Early Release Observations programme; the in-orbit $R^2$ were measured from VIS frames during telescope alignment. 
    }             
    \centering          
    \begin{tabular}{l c c c}    
        \hline\hline
        \noalign{\vskip 3pt}
        Variable & STOP & In-orbit & Required \\
        \noalign{\smallskip}
        \hline 
        \noalign{\smallskip}
        FWHM [\si{\arcsecond}] & 0.146 & 0.141$\pm 0.006$\,$^*$ & 0.155 \\
        Ellipticity $|e|$      & 0.074 & 0.024 (0.05)$^{**}$ & 0.14  \\ 
        $R^2$ [arcsec$^2$]     & 0.053 & 0.048 (0.050)\,$^{**}$ & 0.055 \\              
        $\delta \epsilon$      & $1.5\times 10^{-3}$ & $-$ & $2\times 10^{-3}$ \\
        $\delta$$R^2$/$R^2$    & $1.5\times 10^{-3}$ & $-$ & $2\times 10^{-3}$ \\
        \noalign{\smallskip}
        \hline
    \end{tabular}
    \tablefoot{\\
    $*$ uncertainty range is based on the minimum and maximum measured values in the field.\\
    $**$ average value, the maximum value in the field is given between parentheses.}
\label{tablePerf}      
\end{table}

\subsection{\label{sec:FAR}Performance synthesis before the launch}

Table \ref{tablePerf} shows the performance compared to the requirements. Each performance value is the worst case found among all cases studied, among all field points and all time instants. 

The results of the prelaunch STOP analysis showed that the design is compliant with the applicable requirements under the most extreme conditions, both steady-state and transient. The margins are generally large, even more so taking into account that we have exercised instances where the performance comes close to the specified upper boundary, and the transients are the largest allowed by the design limits.

The goal of the STOP analysis was to study in the design stage the impact of the environment on the PLM to identify and eliminate any significant harmful effects. The analysis was not intended to predict the detailed behaviour of the satellite and payload in-orbit, a purpose for which other tools are available; see Appendix\,\ref{appendix:sec:design-summary} for further details on the thermal design. Nevertheless, the findings of the STOP analysis are useful for the interpretation of the telemetry and the assessment of the in-orbit performance, the subject of the following sections.

\begin{figure*}[ht]
    \centering
    \includegraphics[width=0.85\textwidth]{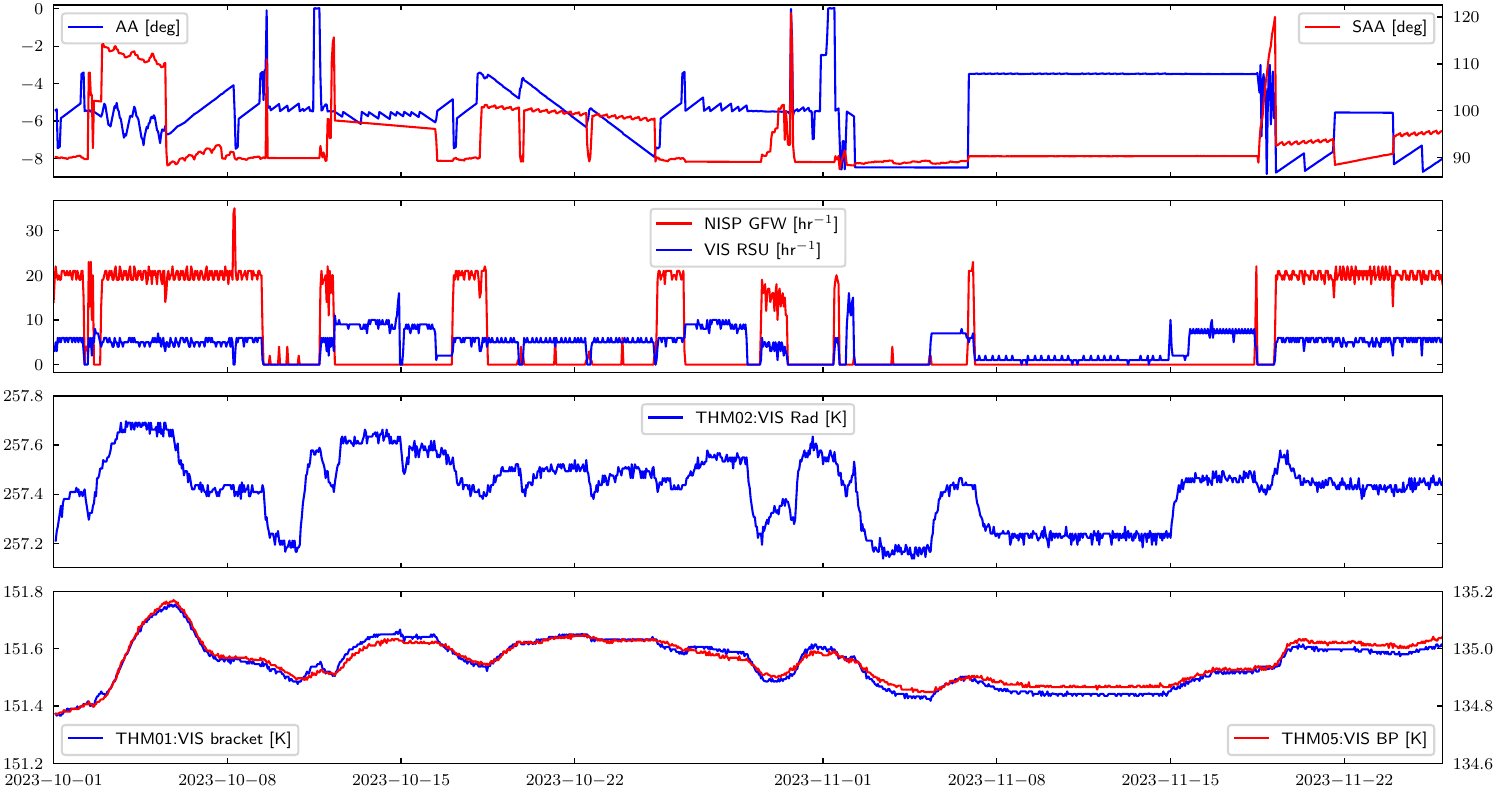}
    \caption{Time histories of spacecraft attitude and key thermal performance parameters obtained from spacecraft telemetry during the \Euclid performance verification phase in the period from 2023-10-01 00:00:00 UTC until 2023-11-26 00:00:00 UTC. {\bf Upper panel:} the variations in SAA and AA. Dedicated rotations of the SAA (red line) and AA (blue line) were performed to verify the thermal response of the system. {\bf Upper middle panel:} the history of the cadence (in cycles per hour) of the NISP grism and filter wheel (GFW, red line), and the VIS shutter (RSU, blue line). {\bf Lower middle panel:} the history of the VIS radiator temperature registered by sensor THM02 (blue line), which is an indicator of the VIS instrument activity by means of its power dissipation. {\bf Lower panel:} the temperature responses of a baseplate sensor placed near the VIS bracket (THM05, red line) and a sensor mounted directly on the VIS \ac{FPA} bracket (THM01, blue line).}
\label{fig:temp-history-2023}
\end{figure*}

\section{\label{sec:commissioning}STOP in-orbit}
\subsection{In-orbit situation}
 
After launch on 1 July 2023, the STOP model was used in support of the commissioning and performance verification phases, to help the interpretation of the in-orbit performance and to verify the STOP analysis predictions.

We examined the spacecraft attitude relative to the Sun, \ac{AA} and \ac{SAA}, and data from \ac{VIS} and \ac{NISP} units whose thermal dissipation affects the telescope. These data were extracted from satellite housekeeping telemetry. Temperature sensors are mounted at thermally important positions in the \ac{PLM} providing means to monitor thermal responses to environmental changes. In particular, 20 high-resolution (<\,10\,mK) thermistors, complemented by a number of lower-resolution (<\,60\,mK) sensors, monitor the thermal state of the baseplate. The record shows that, while the mean temperature is different at each position, the temperature variations are nearly the same, so that any one of the sensors may be used as an index of the thermal state of the baseplate, which correlates with the \ac{IQ}. Variations in telescope structure and image quality cannot be monitored from housekeeping telemetry but must be inferred from stellar images obtained with \ac{VIS}.

Instrument activities generating heat dissipation were deduced from the \ac{RSU} movements for \ac{VIS}, and the filter and grism wheel movements of \ac{NISP}. These movements serve as a close proxy for the dissipation cycle of the instruments. The effect on temperature can be detected only when there is a long sequence (> 12 h) of identical cycles that add up. VIS bias or dark frames, which are read out without moving the \ac{RSU}, cannot be detected in the temperature curve unless they occur in a long sequence during dedicated calibration operations. For the \ac{VIS} effect on the baseplate, we cannot distinguish between heating by the shutter motor or heating by the \ac{ROE} via the flex wires connecting the \ac{ROE} to the \ac{FPA}. 

The spacecraft operating attitude domain was redefined after a parasitic straylight component was found in the \ac{VIS} images at positive \ac{AA}, see \cite{EuclidSkyOverview}. Subsequent analysis showed that the MPS boom is illuminated at AA > $-$2\degf5 deg. The limiting angles used for the construction of the survey had to be set to 87\degf1 < \ac{SAA}  < 120\degf9 and  $-$8\degf4 < AA < $-$3\degf0 to meet the survey requirements while keeping the straylight at an acceptable low level. The negative azimuth limit is 0\degf4 beyond the limit that was considered for the STOP analysis.

During spacecraft commissioning it was found that switching on/off by the telemetry radio frequency transmitter in the \ac{SVM} induced an unwanted rotation around the spacecraft $Z$ axis, causing a disturbance in the pointing. It was decided to permanently leave the transmitter on to minimise disturbance \citep{Gottero2024}. This was done on 2023-10-06, raising the average $T_\mathrm{BP}$ with a few tens of mK, but below 50 mK.

\subsection{In-orbit thermal performance}
\label{sub:flight-thermal-performance}
During the \Euclid performance verification phase from October to early December 2023, numerous calibration activities were carried out, requiring different operating modes from those used during the survey. These activities caused exceptional environmental stresses on the telescope and provided extreme situations suitable for studying thermo-optical effects. 

Before launch, only the Sun incidence angles SAA and AA were considered to be the dominant sources of temperature variation. The in-orbit data indicated that non-standard (i.e. non-survey type) instrument operations unexpectedly dominated the baseplate temperature history. Taking the time histories of various observables, we shall first assess the impact of attitude changes (\cref{subsub:flight-impact-angles}) and subsequently investigate the additional impact of other causes of temperature changes (\cref{subsub:flight-impact-instruments}).

\begin{figure*}[ht]
    \centering
    \includegraphics[width=0.85\textwidth]{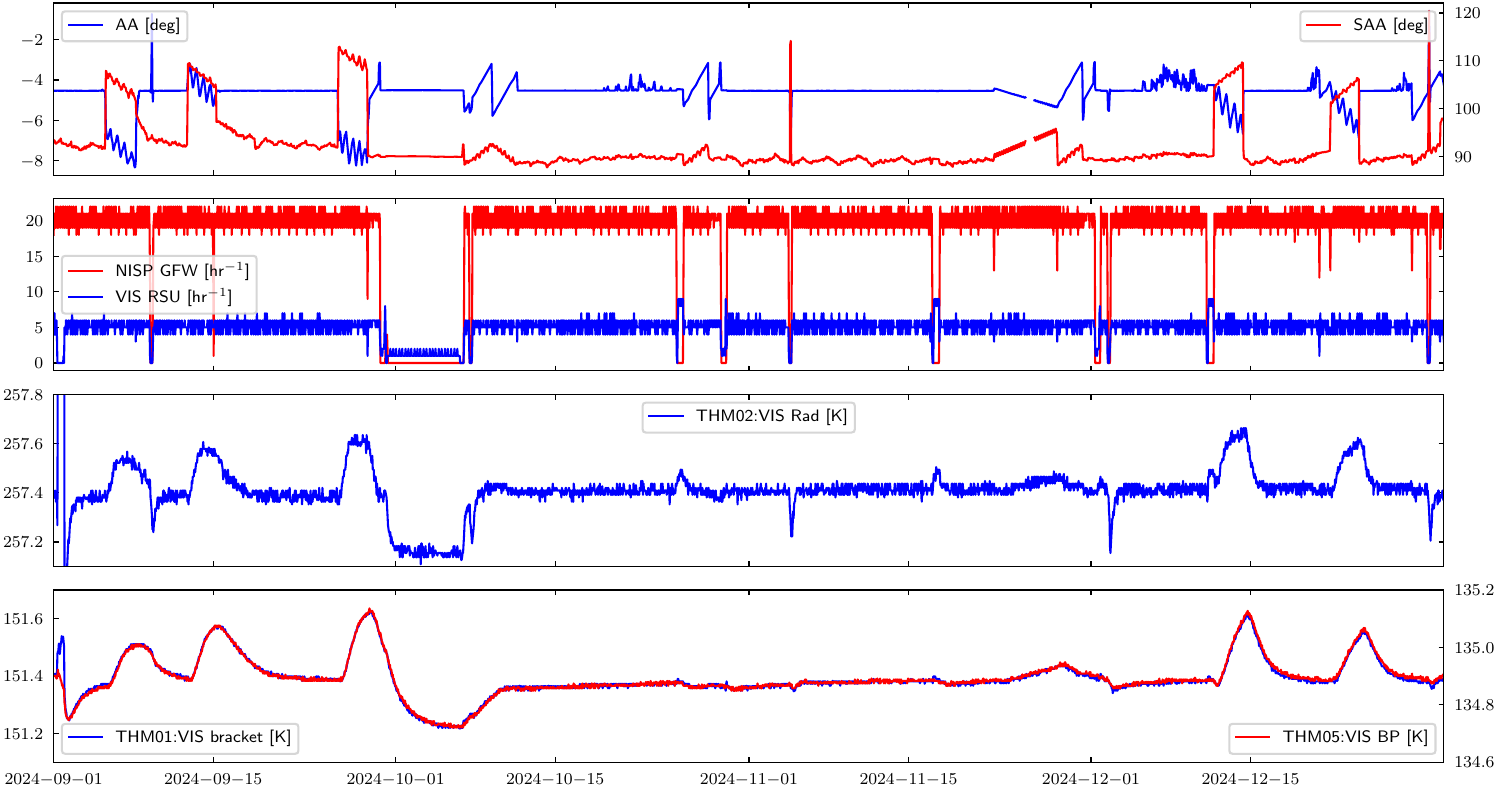}
    \caption{Time histories of spacecraft attitude and thermal performance parameters obtained from spacecraft telemetry during the period from 2024-09-01 00:00:00 UTC until 2025-01-01 00:00:00 UTC. 
    {\bf Upper panel:} the variations in SAA and AA.
    {\bf Upper middle panel:} the history of the cadence (in cycles per hour) of the NISP GFW (red line), and the VIS \ac{RSU} (blue line). 
    {\bf Lower middle panel:} the history of the VIS radiator temperature from sensor THM02.
    {\bf Lower panel:} temperature responses of baseplate sensor placed near the VIS bracket THM05 (red line) and a sensor THM01 directly on the VIS \ac{FPA} bracket (blue line).
    }
\label{fig:temp-history-2024}
\end{figure*}

The time histories of \ac{AA} and \ac{SAA}, and corresponding temperatures of the baseplate, \ac{VIS} \ac{FPA}, and the \ac{VIS} radiator are shown in \cref{fig:temp-history-2023}, covering the period 2023-10-01 until 2023-11-26, and \cref{fig:temp-history-2024}, covering the period 2024-09-01 until 2024-12-31. The baseplate temperature $T_\mathrm{BP}$ was obtained from  sensor THM05 located  close to the \ac{VIS} instrument. The \ac{VIS} \ac{FPA} temperature was obtained from sensor THM01 located on the \ac{VIS} \ac{FPA} bracket, while sensor THM02 provided the \ac{VIS} radiator temperature.

\subsubsection{Thermal impact of SAA and AA variations} \label{subsub:flight-impact-angles}

\cref{fig:temp-history-2023} includes four episodes in which the Solar aspect angle was moved high above 90\degree\, and remained there for several days. In the first such episode, the ``hot case'', a step up of \ac{SAA} from 90\degree\, to 113\degree\, was executed on 2023-10-02, followed by a 3 day stay on an average of 110\degree, and then a step down back to  90\degree\, on 2023-10-05. The step up was accompanied by an exponential increase of $T_\mathrm{BP}$ (THM05), headed for a new equilibrium about 400\,mK above the initial state, to be reached after about 4 days, but interrupted after 3 days by an \ac{SAA} step down back to 90\degree, initiating an exponential drop levelling $\approx$ 200\,mK below the peak. 

Three more \ac{SAA} steps, of smaller magnitude, were then executed:  from 90\degree\, to 97\degree\, starting on 13 October and lasting about 4 days; from 90\degree\, to 100\degree\, average starting on 2023-10-18 and lasting about 7 days; and from 90\degree\, to 99\degree\, average around 2023-10-30 lasting less than 1 day. The longer steps with around 10\degree\, \ac{SAA} variation cause the baseplate temperature to rise by about 100\,mK, always followed by an exponential drop when the \ac{SAA} is brought back to 90\degree.

All episodes of \ac{SAA} rise exhibited an exponential time constant of approximately 1 day and a temperature rise rate, from start to plateau, of 5 to 15\,mK per degree of SAA. These numbers are rough estimates deduced after accounting for other phenomena acting on the temperature, described below.

From 2023-10-30 on, the \ac{SAA} remained around 90\degree, while the baseplate temperature underwent a slow descent, the tail of the exponential drop after the last step down, modulated by other effects that cannot be explained by the \ac{SAA}.

A thermal response to changes in AA would be expected in \cref{fig:temp-history-2023}, from 2023-10-18 to 2023-10-25 (sloping in steps from $-3{\fdg}8$\,  to $-$8\degree\, over 7 days) and from 2023-11-07 (large step from $-8{\fdg}5$\, to $-3{\fdg}5$). However, the effect of AA on $T_{\rm BP}$ is hardly discernible. This is most clearly seen at the time of the large AA step on 2023-11-07. The temperature began to rise some time earlier, therefore the AA step cannot be the cause of the rise. The temperature started to drop soon after the step, and the positive AA step could be the cause of the temperature drop. However, the earlier AA-sloping segment contradicts this hypothesis: a negative AA trend should have caused a temperature rise, which did not appear.

Neither \ac{SAA}  nor AA changes can account for numerous small features of the $T_\mathrm{BP}$ curve of \cref{fig:temp-history-2023}, addressed in the next section. 

\subsubsection{Dependence on instrument activities} \label{subsub:flight-impact-instruments}

The temperature effects of the \ac{VIS} operation are evident in the timeline of the \ac{VIS} radiator (THM02) shown in \cref{fig:temp-history-2023}. The radiator enables the cooling of the \ac{VIS} proximity electronics mounted behind the \ac{FPA} \citep{EuclidSkyVIS}. The radiator temperature, which depends on \ac{SAA} like all other elements, rises quickly to a new equilibrium when the \ac{VIS} exposure cadence changes to a higher value and drops with changes to a lower cadence. The temperature of the \ac{VIS} FPA bracket (THM01) follows the \ac{VIS} radiator, with a damped magnitude and a longer time constant close to that of the baseplate temperature (THM05). At times when the radiator temperature changes abruptly, indicating a change in the cadence of the readouts, the temperature fluctuation of THM01 gets slightly larger or smaller than that of THM05, causing the two curves in \cref{fig:temp-history-2023} to separate.

\Cref{fig:temp-history-2023} covers parts of the \ac{PV} phase and, in particular, the \ac{PDC} demonstration between 2023-11-01 and 2023-11-27.  During this period, numerous calibration activities were performed, subjecting the PLM to a rapid succession of enhanced thermal perturbations unlike anything experienced during the normal survey. It is exceptionally suitable for assessing the impact of payload operations on the thermal stability of the PLM. Here, the temperature curves show, in addition to the global impact of the \ac{SAA}, a distinct signature of each calibration block.

The temperature effect of the \ac{VIS} operation can be appreciated on 2023-11-05. At that epoch, the \ac{SAA} had been stable around 90\degree\, for 5.5 days, and the \ac{VIS} and NISP instruments had been inactive for 4 days; as a consequence, both the \ac{VIS} radiator and the baseplate had dropped to their lowest temperature.  On 2023-11-05 08:19:25 UTC, a series of 256 short (92\,s) \ac{VIS} exposures was initiated, lasting 1.5 days. The high cadence was associated with a temperature increase of the \ac{VIS} radiator by 300 mK, and of the baseplate by 50\,mK. The phenomenon can be explained by each readout event injecting heat pulses by the \ac{RSU} motor and the readout electronics, via the dissipation directly to the baseplate and via the thermal straps connecting FPA bracket and baseplate, respectively.

Late on 2023-11-06, AA was moved from $-8\fdg5$ to $-3\fdg5$, where it stayed for about 6 days, during which a series of long exposures of 560\,s was taken. In this period, we observed a drop of 200\,mK in the \ac{VIS} radiator temperature, and an exponential drop in $T_\mathrm{BP}$ of 35\,mK, with a time constant of 29 hours (the time constant of the baseplate was measured during the telescope cool-down of 5 days from July 15 and was found to be 20 hours, while that of M1 in the same period was about 33 hours). As explained above, this temperature drop can be completely attributed to the lower dissipation due to the longer time intervals between successive readouts, and not to the change of \ac{AA}.

A small impact of the \ac{NISP} filter and grism wheel mechanism motors on $T_\mathrm{BP}$ can be inferred when comparing temperatures at epochs when the wheels are on and off. We deduce a contribution of the NISP wheels of less than 30 mK. 

\subsubsection{Thermal stability during nominal survey operations}
\label{subsub:thermal-during-survey}

The thermal history of the system for a four-month period during the nominal mission is presented in \cref{fig:temp-history-2024} displaying the same parameters as in \cref{fig:temp-history-2023}. The baseplate temperature $T_\mathrm{BP}$ registered by THM05 appears to be dominated by five large excursions in \ac{SAA}, which for each case was rotated from around 90\degree, to more than 105\degree, and back, causing $T_\mathrm{BP}$ to increase by more than 100 mK. The large \ac{SAA} rotations were triggered by observations of several days in the \Euclid Deep Field South for dedicated calibrations.

The period between 2024-09-30 to 2024-10-07 was allocated to a biannual calibration block which contains a NISP non-linearity calibration of 6.7 days, which did not require filter wheel movements and was accompanied by low VIS activity. These operations caused a prominent dip in $T_{\rm BP}$ of more than 100 mK below the baseline. Interruptions involving large \ac{SAA} changes confined during the 12\,h monthly satellite maintenance operations had a much smaller effect on $T_\mathrm{BP}$. Other, less prominent temperature excursions can all be associated with calibration observations causing changes in \ac{SAA}, or a reduced VIS cadence in case of recurrent \ac{PSF} calibration observations. The time history shows that the baseplate temperature remained nearly half of the time in transient periods, during the months September 2024 and December 2024, and the first 10 days of October 2024. This is predominantly caused by a few large \ac{SAA} rotations, which were less than two weeks apart in time.

\begin{figure}
    \centering
    \includegraphics[width=0.70\columnwidth]{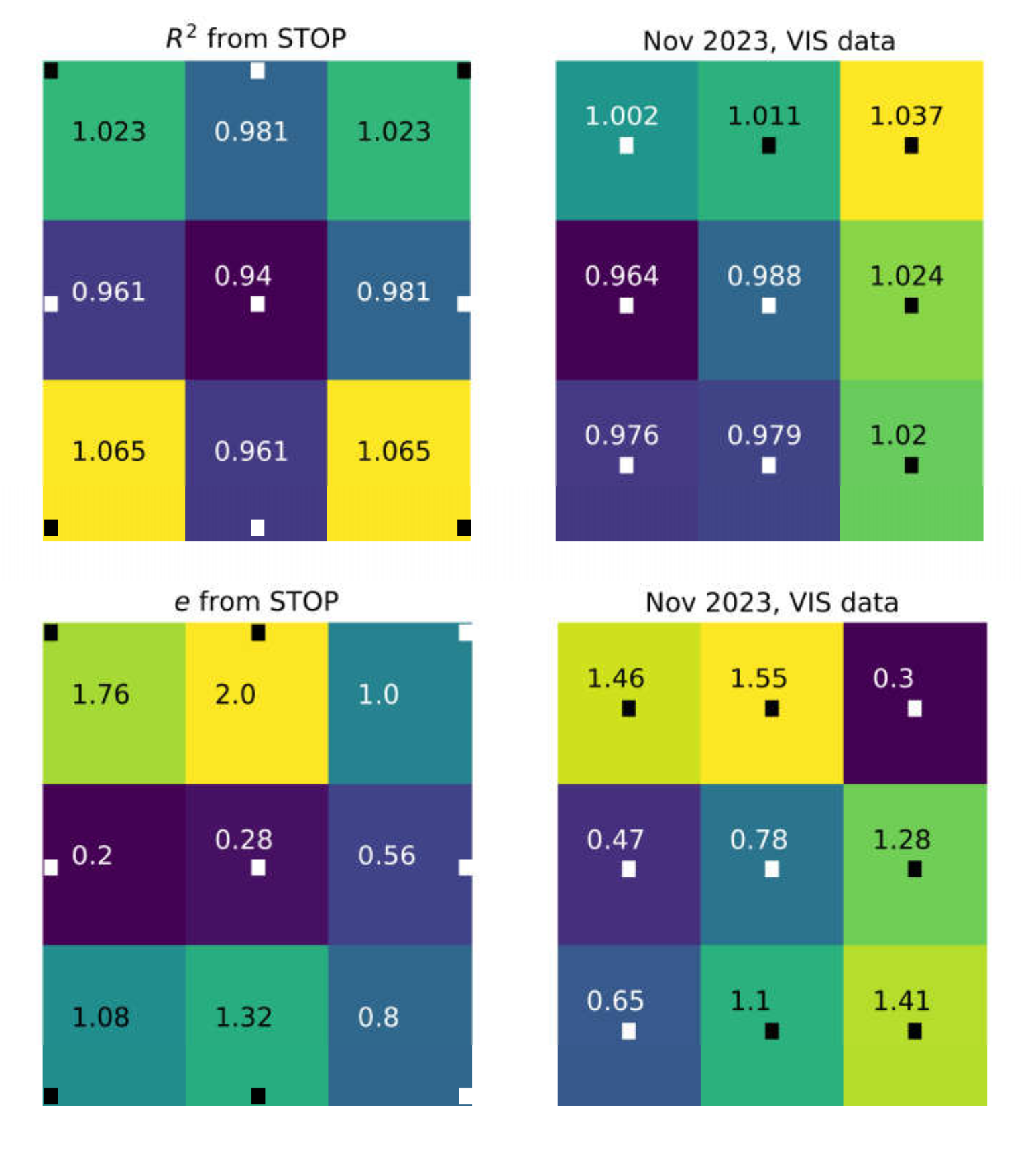}
    \caption{Visualisation of the relative variation of $R^2$ and $e$ over the \ac{VIS} focal plane, $+$X to the right and $+$Y going up. The values are normalised by the average of the 9 values, with average $R^2$ of 0.0479 and 0.0199 arcsec$^2$ and average $e$ of 0.025 and 0.021, for the STOP and in-orbit data, respectively. The difference between the values are due to the usage of a different Gaussian weighting in the calculation of the moments (see \cref{subsub:static-IQ}). The squares mark the reference positions on the FPA where the values have been obtained. {\bf Left diagrams:} the relative $R^2$ (top) and $e$ (bottom) predicted by the STOP analysis. {\bf Right diagrams:} relative $R^2$ and $e$ measured in-orbit in November 2023.}
\label{fig:r2e_FPA_variations}
\end{figure}

\subsection{Image quality}

\subsubsection{Image quality after first alignment of the telescope}
\label{subsub:first-alignment}
Soon after decontamination and cool-down of the \ac{PLM} in July 2023, the telescope was aligned by adjusting the \ac{M2} mechanism which was previously set in launch condition. The whole alignment process was based on images obtained by the \ac{VIS} instrument. Its objective was to optimise the $R^2$ image quality metric over the whole \ac{VIS} \ac{FPA}, with \ac{SAA}=\,90\degree\, and \ac{AA} =\,0\degree. A preliminary corrective motion of the \ac{M2} was carried out on the basis of the \ac{PSF} shape of stars visible on previously acquired \ac{VIS} images. This initial motion brought the \ac{M2} closer to its optimum position. Then, the three degrees of freedom (translation along optical axis, tip, and tilt) of the \ac{M2} mechanism have been scanned successively. Along each scan, acquisitions were performed, and the $R^2$ of the acquired stars spread over the whole field were computed. After each scan, the \ac{M2} position minimising the worst $R^2$ was computed and applied. After alignment, the CODE V optical model was updated to match the in-orbit measurements. The numerical simulations after updating the model showed the excellent optical quality of the telescope, well within the required performance: a worst case $R^2 =0.051$\,arcsec$^2$ and an average $R^2=0.047$\,arcsec$^2$, over the nine reference field points, for a monochromatic source at 800 nm, to be compared with the required value of 0.055 arcsec$^2$ at spacecraft level (\cref{tablePerf}), apportioned to 0.053 arcsec$^2$ at telescope level. A direct extraction of the telescope $R^2$ from the in-orbit acquisitions during the alignment process was also performed. The extraction includes unaccounted contributions from the colours of the stars and the detector \ac{PSF}, which increase the value of $R^2$. However, the result of this analysis was in line with the numerical simulations: a maximum $R^2=0.050$\,arcsec$^2$ and an average $R^2=0.048$\,arcsec$^2$, both within the requirements.

We determined the \ac{FWHM} and ellipticity from a nominal science observation obtained in early September 2023 as part of the \Euclid \ac{ERO} programme \citep{EROData}. The observation of the dwarf irregular galaxy IC10 was selected because it is situated in a field at a low Galactic latitude ($b=3\degree$) with high stellar density. The average \ac{FWHM} and ellipticity $|e|$ measured from a single exposure are listed in \cref{tablePerf}. Despite the additional uncertainties due to the colours of the stars and the detector effects, the in-orbit \ac{IQ} performance appears to be well within the limits indicated by the STOP predictions and requirements.

\subsubsection{Static IQ performance}
\label{subsub:static-IQ}

We obtained the in-orbit \ac{IQ} parameters $R^2$, $e_1$, and $e_2$ by measuring the quadrupole moments of selected stars detected in \ac{VIS} frames, and by excluding the \ac{VIS} short exposures of 95\,s. To overcome an unacceptable high noise caused by cosmic rays, we reduced the width of the Gaussian weighting of $\sigma$ = 0$\farcs$75 (see \cref{sec:requirements})  to $\sigma$ = 0$\farcs$25. This suppression of the wings of the \ac{PSF} produced a systematically smaller value of $R^2$. The resulting values for $R^2$ and ellipticity $e$ cannot be directly compared to the required values but can be used as diagnostics for \ac{IQ} variations. Details of the \ac{IQ} determination are given in Appendix\,\ref{appendix:IQ-on-orbit}.

We investigated the static \ac{IQ} performance in the period from 2023-11-01 to 2023-11-26 by averaging the $R^2$ and $e$ over time intervals of 12\,h for the stars detected in 9 areas of the \ac{VIS} \ac{FPA}. The spatial variation of the mean \ac{IQ} parameters across the FPA is significantly larger than the variation in each area during the measurement period. The relative \ac{IQ} variations in the \ac{VIS} \ac{FPA} are visualised in \cref{fig:r2e_FPA_variations} where the observed results are compared with the \ac{STOP} predictions. The predicted \ac{IQ} values were calculated in the reference field points defined in \cref{FigIMA} where 8 of them are located at the edges of the \ac{VIS} \ac{FPA} to determine extreme design results. This is different from the central positions of the 9 areas (indicated in \cref{fig:r2e_FPA_variations}) which were obtained by spatial averaging of the \ac{IQ} data. The in-orbit $R^2$ variation does not resemble the predicted distribution with a minimum $R^2$ at the centre of the FPA. The in-orbit variation in ellipticity shows better resemblance to the predicted distribution and the magnitude of the relative variation is similar.

\subsubsection{Transient performance: IQ time histories}
\label{subsub:IQ-histories}

\begin{figure}
    \centering
    \includegraphics[width=0.99\columnwidth]{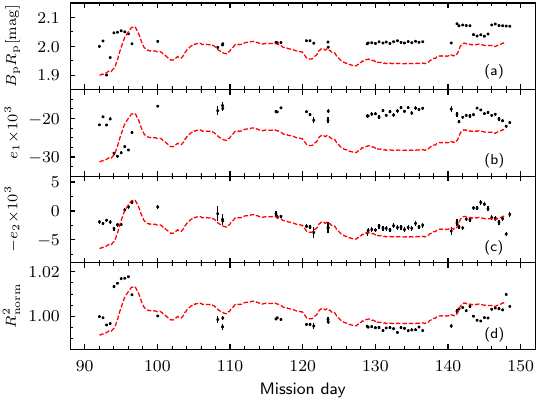}
    \caption{Time history of the observed \ac{IQ} metrics for the period shown in \cref{fig:temp-history-2023}, before the start of the nominal mission, from 2023-10-01 (mission day 92) until 2023-11-26 (mission day 148).{\bf  (a):} the variation of the average {\it Gaia} colour $B_{\rm p}R_{\rm p}$ per 12\,h bin. {\bf (b):} the variation of the ellipticity component $e_1$. {\bf (c):} the variation of the ellipticity component $e_2$. To underscore the correlation with baseplate temperature $-e_2$ is displayed. {\bf (d):} the variation of $R^2$. The red dashed lines indicate baseplate temperature $T_\mathrm{BP}$ (THM05) with arbitrary scaling for reference.}
\label{fig:IQ_history_2023}
\end{figure}

\begin{figure}
    \centering
    \includegraphics[width=0.99\columnwidth]{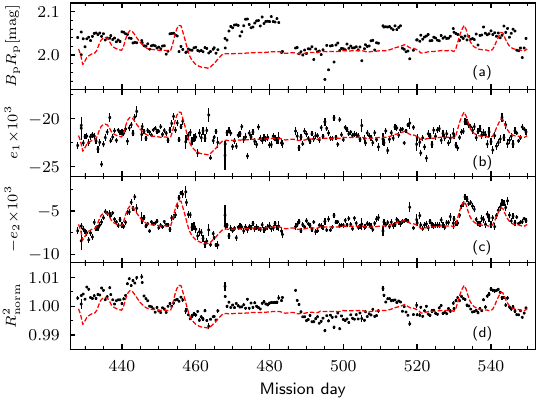}
    \caption{Similar as \cref{fig:IQ_history_2023} but for the \ac{IQ} metrics observed in the period shown in \cref{fig:temp-history-2024}, covering 4 months of nominal mission, from 2024-09-01 (mission day 428) until 2024-12-31 (mission day 549).} 
\label{fig:IQ_history_2024}
\end{figure}

The time histories of $R^2$, ellipticity components $e_1$ and $e_2$, as well as the stellar colours averaged over 12\,h time bins over the periods addressed in \cref{sub:flight-thermal-performance} are presented in Figs.\,\ref{fig:IQ_history_2023} and \ref{fig:IQ_history_2024}.
The selection of stars based on limited colour and magnitude suitable to derive the \ac{IQ} parameters provided sufficient statistics only for time intervals of 12\,h, shorter time intervals did not improve time resolution. The values of $R^2_{\rm norm}$ were obtained by normalising the timelines per \ac{FPA} area by dividing the values by their mean value over the duration of the period. Subsequently, we derived the mean normalised $R^2$ of the 9 \ac{FPA} areas weighted by the variances per time bin, see Appendix\,\ref{appendix:IQ-on-orbit} for details. The values for $e_1$ and $e_2$ are the average of the 9 \ac{FPA} areas without further normalisation. The stellar colours serve as an indicator for possible colour dependencies on the \ac{IQ} metrics. We plotted $-e_2$ to show the correlation better with temperature $T_{\rm BP}$.

$R^2_{\rm norm}$ exhibits a peak-to-peak variation of less than $\approx 2.2\%$ in both time histories. Variations in ellipticity components are substantially greater, up to a factor of 1.7 for $e_1$ and 3.6 for $e_2$ in the 2024 period. The ellipticity component $e_1$ is more than 3 times larger than $e_2$, making it the dominant component of $e$. During the 2023 period, $e_2$ changed sign, indicating that $e_2$ can be zero during periods of time. In addition, the mean values of the ellipticity components over the two time periods are different, which could be due to differences between pre-launch and nominal survey conditions. Despite the observed variations, the FPA-averaged $R^2$ and $|e|$ remain within the required values assuming $R^2 = R^2_{\rm norm}{\times}R^2_{\rm in-orbit}$, where $R_{\rm in-orbit}$ is the in-orbit value provided in \cref{tablePerf}.

The time history of the 2024 period (\cref{fig:IQ_history_2024}) allows us to assess the \ac{IQ} metrics during the routine operations over a long time interval without gaps in the VIS observations as large as seen in \cref{fig:IQ_history_2023}. $T_{\rm BP}$ exhibits 5 prominent peaks, which all can be associated with observations of the southern calibration field for spectroscopy (CPC-South), requiring large \ac{SAA} rotations affecting $T_{\rm BP}$. The variations of $R^2_{\rm norm}$ in \cref{fig:IQ_history_2024}d roughly match those seen in $T_{\rm BP}$, with higher $R^2_{\rm norm}$ when $T_\mathrm{BP}$ is higher. Four of the five prominent peaks in $T_{\rm BP}$ are accompanied by peaks in $R^2_{\rm norm}$. However, higher values of $R^2_{\rm norm}$ can also be associated with survey periods in which the stars are redder. During mission days 467 to 482 \Euclid observed a region in the Deep Field North followed by a nearby survey patch at low Galactic latitude (close to $b=\,+$30\degree) where the selected stars are on average $\approx0.06\,{\rm mag}$ redder per bin than during the earlier and later observations. The period 510--515 can be associated with observations of the COSMOS calibration field, where the selected stars are $\approx 0.04\,{\rm mag}$ redder. During these periods $R^2_{\rm norm}$ is higher by 0.3--0.5\%. In addition, the second and third temperature peaks, occurring on mission days $\approx$\,442 and $\approx$\,456, are accompanied by redder $B_{\rm p}R_{\rm p}$ of 0.02 and 0.03 mag, respectively. We infer that our preselection of stars in a limited colour range does not prevent a bias in the values of $R^2_{\rm norm}$ due to observations of survey patches containing significantly redder stars with respect to the observations taken before and after. The colour variations could be reduced by a narrower range in $B_{\rm p}R_{\rm p}$ for the selection of stars, but then the statistical scatter in the \ac{IQ} timelines would become too large. Fortunately, the temperature peaks detected in December 2024 on mission days ${\approx}$\,533 and ${\approx}$\,542 are not accompanied by significant colour excursions. Therefore, we attribute the peaks in $R^2_{\rm norm}$ to variations in temperature $T_{\rm BP}$.

The time history of $e_2$ in \cref{fig:IQ_history_2024}c shows a striking correlation with $T_\mathrm{BP}$ and no colour dependence. The ellipticity component $-e_2$ traces all the peaks in $T_{\rm BP}$ with the same relative amplitudes. It also closely follows the course of the baseline of $T_{\rm BP}$. The correlation is less tight in \cref{fig:IQ_history_2023}c, and there might be an indication that $e_2$ is affected by the colour of the stars; see mission days 144 and 145. For $e_1$, we observe that the strongest temperature peaks are accompanied by 10--15\% higher values of $e_1$, with no obvious colour dependency. However,  \cref{fig:IQ_history_2024}b suggests a weaker correlation with $T_\mathrm{BP}$ than was observed for $R^2_{\rm norm}$ and $e_2$.

\subsubsection{Other IQ metrics: trefoil}
\label{subsub:IQ-trefoil}

In addition to the quadrupole moments metrics, other third-order metrics were also visible, such as the trefoil, producing a ``triangular'' shaped \ac{PSF}. A degree of trefoil was predicted as a result of the introduction of a special mechanical device to correct for a slight astigmatism by M1 during telescope development. The updated CODE V model tuned for sky observations predicts a \ac{WFE} contribution of the trefoil Zernike terms ($z_{10}$ and $z_{11}$) of less than 14\,nm rms in the worst position. This value is negligible in the \ac{WFE} budget ($\approx$\,50\,nm) required to be well within the \ac{IQ} performance in the 9 reference field points. We did not analyse the on-orbit dependency of the trefoil with variations in $T_{\rm BP}$, we refer to Whittam et al. (in preparation)  for a detailed investigation.

\begin{table*}
    \caption{Estimated correlations based on the analysis of the in-orbit data. In-orbit $R^2$ and $e$ were extracted from data obtained in the period September to December 2024, averaged over intervals of 12 h. Temperature sensor THM05 was used for $T_{\rm BP}$.}             
    \centering 
    \begin{tabular}{l c c l }    
        \hline\hline
        \noalign{\vskip 3pt}
        relation & in orbit & STOP prediction & comment \\
        \noalign{\smallskip}
        \hline 
        \noalign{\smallskip}
        ${\Delta} T_{\rm BP}/{\Delta}$SAA (mK\,deg\inv) & 5--15 & 2.6 & SAA >90\degree ~for STOP \\
        ${\Delta}R^2_{\rm norm}/{\Delta}T_{\rm BP}$ (\%\,K\inv) & 3--6   & 3.7 & in-orbit value from peak on mission day 533\\ 
        &  &  & estimated 30\% uncertainty in $R^2$ due to colour\\
        ${\Delta}R^2_{\rm norm}/{\Delta}$SAA (\%\,{\rm deg}\inv) &  0.03--0.05  & 0.01 & SAA >90\degree; in-orbit peak on mission day 533\\
        &  &  & estimated 30\% uncertainty in $R^2$ due to colour\\
        ${\Delta}e_{\rm norm}/{\Delta}T_{\rm BP}$ (\%\,K\inv) &  50--70 & 70& variation of mean normalised ellipticity |$e$|\\ 
        ${\Delta}e_1/{\Delta}T_{\rm BP}$ (K\inv) &  0.01 & -& estimated from \cref{fig:IQ_history_2024}b\\ 
        ${\Delta}e_2/{\Delta}T_{\rm BP}$ (K\inv) &  $-$0.013 & -& estimated from \cref{fig:IQ_history_2024}c\\ 
        ${\Delta}e_{\rm norm}/{\Delta}$SAA ($\%\, \deg\inv)$ &  0.5--0.7  & 0.2 & SAA >90\degree ~for STOP; \\
        Baseplate thermal time constant~(h) &  20 & 12--15 & the larger STOP value was derived from\\
        &  &  & system parameters updated close to launch\\
        M1 thermal time constant~(h) &  33 & 33--35 & \\
        \noalign{\smallskip}
        \hline                  
    \end{tabular}
\label{tab:on-orbit_correlations} 
\end{table*}

\section{\label{sec:discussion}Discussion}

\subsection{Synthesis of thermal and structural features from the STOP analysis}

Thermal analysis revealed the importance of the baseplate as the thermal regulator of the telescope. We note that PLM heaters are not used for the regulation of temperature variations but are controlled by open-loop settings to maintain its temperature constant. The temperatures of the main optical elements show a linear dependency with the average temperature of the baseplate, suggesting that the dominant temperature-driven telescope aberration is a global effect and not due to local distortions in the optical train or by individual mirrors. These findings confirm the purpose of the all-SiC homothetic design, where the entire telescope structure expands and contracts homogeneously with temperature. However, the linear dependence with the baseplate temperature is not uniform, indicating that the effect is not completely homothetic still causing optical aberration albeit on a lower level than for a non-homothetic design.

In the analysis, the Sun’s incidence angles, \ac{AA} and \ac{SAA}, are the temperature drivers. Within the stated limits of these angles, the maximum temperature fluctuation was found to be $\approx$\,0.3\,K, while during the 7-year lifetime, the mean temperature was predicted to vary by $\approx$\,1.5\,K. The EOL condition assumes a 7-year degradation of the surfaces responsible for thermal regulation, as well as the extremes of the seasonal variation. Hence, if all of the variation is attributed to a gradual degradation process, the STOP analysis predicts a continuous increase of the baseplate temperature of $\approx$\,200\,mK/year. Temperature variation at this level can be monitored with sufficient accuracy since the temperature sensors have a resolution better than 10\,mK.

Transient analysis revealed that the telescope temperature stability on the 11\,000\,s time scale has a long-term and a short-term component. The dominant component, with a long time constant of 12 hours, describes the heating/cooling of the PLM following the variation of the Solar aspect angle and of the internal dissipation. The fast (about 3\,h) component is revealed in the derivative of the temperature variation of the baseplate. We understand the two terms as associated with the different time constants of the heavily insulated all-SiC PLM and the more exposed SVM, which experiences a rapid and large temperature excursion as the Sun moves below the horizontal plane. The fast component was shown to have a counterpart in the displacement of M2 and other mirrors, acting as a damped oscillation, with a quick rise and a slower exponential return. Later optical analysis showed that this phenomenon under extreme conditions approaches the upper limit of the IQ stability boundary.

\subsection{Observed thermal properties in-orbit}

As to the spacecraft attitude, SAA changes affect the baseplate temperature, whereas the effect of AA is hardly noticeable if present at all. A large AA step of 5\degree\, did not provide evidence for a related $T_\mathrm{BP}$ variation (\cref{subsub:flight-impact-angles}). Since the operational range in AA was reduced to 5\degf4, which is only 30\% of the range considered in the STOP analysis, we do not expect significant baseplate temperature changes due to AA variations alone. 

For the SAA, we observe an increase in $T_\mathrm{BP}$ when the SAA increases above 90\degree. For constant ${\Delta}$SAA, the temperature increases approximately following a logistic function, reaching a plateau after about 4 days. Short-duration visits to high SAA of less than 12\,h have little impact. The STOP analysis predicted an effect of SAA on baseplate temperature; however, the magnitude of the effect was 1--2\,\mbox{mK} for ${\Delta}$SAA = 10\degree, while an average of about 5--15\,mK/deg is observed (see \cref{tab:on-orbit_correlations}). This suggests a larger than predicted heat input from the \ac{SVM} to the baseplate, possibly due to a slight overestimation of thermal insulation by the \ac{MLI} devoted to the protection of the \ac{SVM} bottom part against the Sun.

The operation of the instruments causes significant baseplate temperature variations. In the case of \ac{VIS}, a higher cadence of the exposures leads to an increase of $T_\mathrm{BP}$. For \ac{NISP}, filter and grism wheel actuations are a source of power dissipation leading to $T_\mathrm{BP}$ fluctuations at a level of a few 10\,mK. These effects were not studied before launch because the instrument dissipation was assumed constant. After they were detected, the STOP model was employed to verify the heat exchange between the \ac{VIS} instrument components, the \ac{VIS} radiator, and baseplate. To simulate the baseplate temperature variation, we applied a power reduction step to the FPA electronics ($-$0.5\,W) and RSU ($-$0.2\,W), corresponding to the extra power associated to a readout event. The subsequent evolution of the temperatures was monitored for six days. The model provided time constants and temperature variations in qualitative agreement with the orbit data. 

\subsection{Image quality}

The first VIS measurements after the telescope alignment showed that the \ac{IQ} parameters $R^2$, \ac{FWHM}, and ellipticity $e$ are in line with the \ac{STOP} predictions and meet the design requirements; see \cref{tablePerf}. The in-orbit \ac{IQ} values averaged over the VIS \ac{FPA} given in \cref{tablePerf} indicate comfortable margins, without correcting for the spectral energy distribution of the stars (colours) and the known detector effects.

The STOP analysis indicated that the contribution of the structural deformation of the truss to image quality should be small, less than 0.5\% (\cref{subsub:IQ-static}). The main cause of variations is the deformation of the mirrors due to temperature variations. From the analysis of the Zernike terms in the transient case it was found that the telescope aberration is mainly the defocus term (\cref{subsub:IQ-transient}) and is proportional to the variation of the temperature of the baseplate (\cref{sub:thermal-analysis}).

The time histories of the \ac{IQ} metrics were determined from VIS frames obtained during early operations and during a period of four months during the nominal mission of \Euclid. The \ac{IQ} values are not directly comparable with the required values; therefore, relative comparisons were made (\cref{subsub:static-IQ}). The predicted $R^2$ distribution over the FPA is found to be significantly different from the observed distribution (\cref{fig:r2e_FPA_variations}). This suggests that the in-orbit wavefront error differs from the predicted \ac{WFE} which was based on on-ground measurements of the telescope's optical parameters. The observed ellipticity distribution in the FPA shows a better match with the notion that the in-orbit $e$ is dominated by $e_1$. For both \ac{IQ} parameters, the observed relative variations over the FPA are smaller and within the peak-to-peak variations predicted by \ac{STOP} suggesting that the deviations are within the requirements.

The in-orbit variations of the \ac{IQ} parameters $R^2_{\rm norm}$, $e_1$, and $e_2$ exhibit a dependence on $T_\mathrm{BP}$ (Figs.\,\ref{fig:IQ_history_2023} and \ref{fig:IQ_history_2024}). We found that our measurement of $R^2_{\rm norm}$ is biased by the mean colour of the stars in the field, an effect we could not remove by applying a narrower colour selection without losing statistical significance, but we can identify $R^2$ variations without colour bias. The tight correlation between $e_2$ and $T_{\rm BP}$ and to a lesser extent the correlation of $R^2$ and $e_1$ with $T_{\rm BP}$ highlight the role of $T_{\rm BP}$ and the thermo-optical time constants in \ac{IQ}, as predicted by the STOP analysis. The estimated dependencies are given in \cref{tab:on-orbit_correlations}.
Assuming that during the wide survey the instrument activities have a constant cadence, variations in image quality of science data can mainly be attributed to variations in SAA, when considering stars with the same colour. We note that all significant baseplate temperature transients during the nominal survey period investigated by us are triggered by deep-field or calibration observations that require large SAA rotations and observation periods longer than 12\,h. Large SAA excursions with durations shorter than 6\,h, seen during spacecraft maintenance periods, caused no significant \ac{IQ} variations.

We were unable to verify the stability requirements in-orbit to the required stability time resolution of 11\,000\,s due to the statistical noise in our measurements over these time intervals. The minimum time interval considered by us is 12\,h. The observed variations after large SAA steps indicate that the resulting peak values of $R^2$, $e_1$ and $e_2$ are reached after $\approx$\,48 hours. Assuming a pessimistic $\Delta T_{\rm BP}/\Delta {\rm SAA} =$ 15\,mK/deg (\cref{tab:on-orbit_correlations}), we obtain $\Delta {e_1}/\Delta {\rm SAA}= 2\times10^{-4}$. An instantaneous \ac{SAA} step of 15\degree\, would result in a continuous $e_1$ increase of $3\times10^{-3}$ over 48 hours. During that period, the stability science requirement is violated ($2\times10^{-4}$ over 11\,000~s, see \cref{tableReq}), but not the spacecraft stability requirement ($2\times10^{-3}$ over 11\,000~s). For $\Delta R^2 / R^2$, the science requirement over time intervals of 11\,000\,s would also be violated for a slew of 15\degree, but not the spacecraft requirement, which would be violated if the change happened in 24\,h instead of 48\,h. The estimated variations of $e_2$ and $R^2$ could be more severe in some intervals, but we do not have the time resolution. Our estimation shows that the margin in the $R^2$ stability requirement is good to within a factor of 2.

\section{\label{sec:conclusions}Conclusions}

We have presented a mathematical model, which includes thermal, structural, and optical components, to ensure that the design of the \Euclid spacecraft can support stringent image quality requirements to meet its scientific objectives. The model was adapted to the development of the satellite and was refined to predict the in-orbit performance of the final design of the satellite, confirming that the requirements would be fulfilled during the lifetime of the mission. The model was used to investigate the response of the telescope to extremes in the expected thermal environment, also accounting for ageing during the mission. It showed that the defocus is the dominant aberration, and is due to the overall deformation of the optical train and mirror surfaces, with a linear dependency on the temperature of the telescope baseplate.

We compared prelaunch predictions with in-orbit performance data obtained from housekeeping telemetry and VIS exposures. Changes in SAA are confirmed to be the dominant cause of baseplate temperature fluctuations. We observe a temperature increase as large as 400\,mK when SAA moves to 120\degree and remains there for 4 days or more. AA rotations cause no obvious temperature variations as long as they are confined to a smaller range in the negative AA domain, as is the case in the present survey. The SAA dependency is predicted by STOP but with a much smaller amplitude. We attribute the observed behaviour to a possible under-performance of the MLI attached to the bottom of the SVM. However, it should be noted that the deviation from the prediction is extremely small in absolute terms and completely compatible with the uncertainties associated with a thermal analysis based on state-of-the-art mathematical models used in the space industry.

The in-orbit time constants are found to be around 1 day for the baseplate and 33 hours for the primary mirror M1. These values are significantly longer than predicted by the STOP model and are likely due to numerous worst-case assumptions, including less efficient thermal insulation between the top side of the \ac{SVM} and baseplate, adopted to make the design robust.

In addition to changes in solar attitude, activities by the VIS instrument are the second dominant cause of baseplate temperature variations, which can be up to 100\,mK under nominal survey operations. These changes can be associated with the cadence of the readouts. NISP activities, which were monitored by the instrument's filter wheel movements, have a lower impact on the baseplate temperature. We note that time-dependent NISP activities were not considered in the STOP analysis.

Monitoring in-orbit image quality parameters $R^2$ and ellipticity $e$ over two periods during the mission, including a continuous survey period of four months, confirms the dependence of both parameters on the baseplate temperature, as predicted by the STOP model. Although the ellipticity varies according to the prediction and remains within the requirements, the dependence of $R^2$ on \ac{SAA} appears significantly higher. Applied steps in \ac{SAA} of more than $\approx$\,10\degree --20\degree, could cause a violation of the IQ stability requirement for periods of 1 to 2 days, which corresponds to 20--40 survey observations. We were unable to verify the predicted compliance of the stability requirement during shorter periods, due to the statistical noise in our measurement method.

We conclude that the STOP analysis proved useful during the design phase in correctly estimating the observed behaviour, to the extent needed during development. In this paper, we provide the first quantitative assessment of the measured effects of environmental parameters on image quality. The observed in-orbit deviations, while confirming the good design choices, deserve further investigation.

\begin{acknowledgements}
\AckECol
\end{acknowledgements}

\bibliography{Euclid_STOP_descriptionRefs, Euclid}
%

\begin{appendix}
\nolinenumbers

\section{\label{appendix:sec:design-summary}\Euclid design and development summary}

\begin{figure}
    \centering
    \includegraphics[width=0.8\columnwidth]{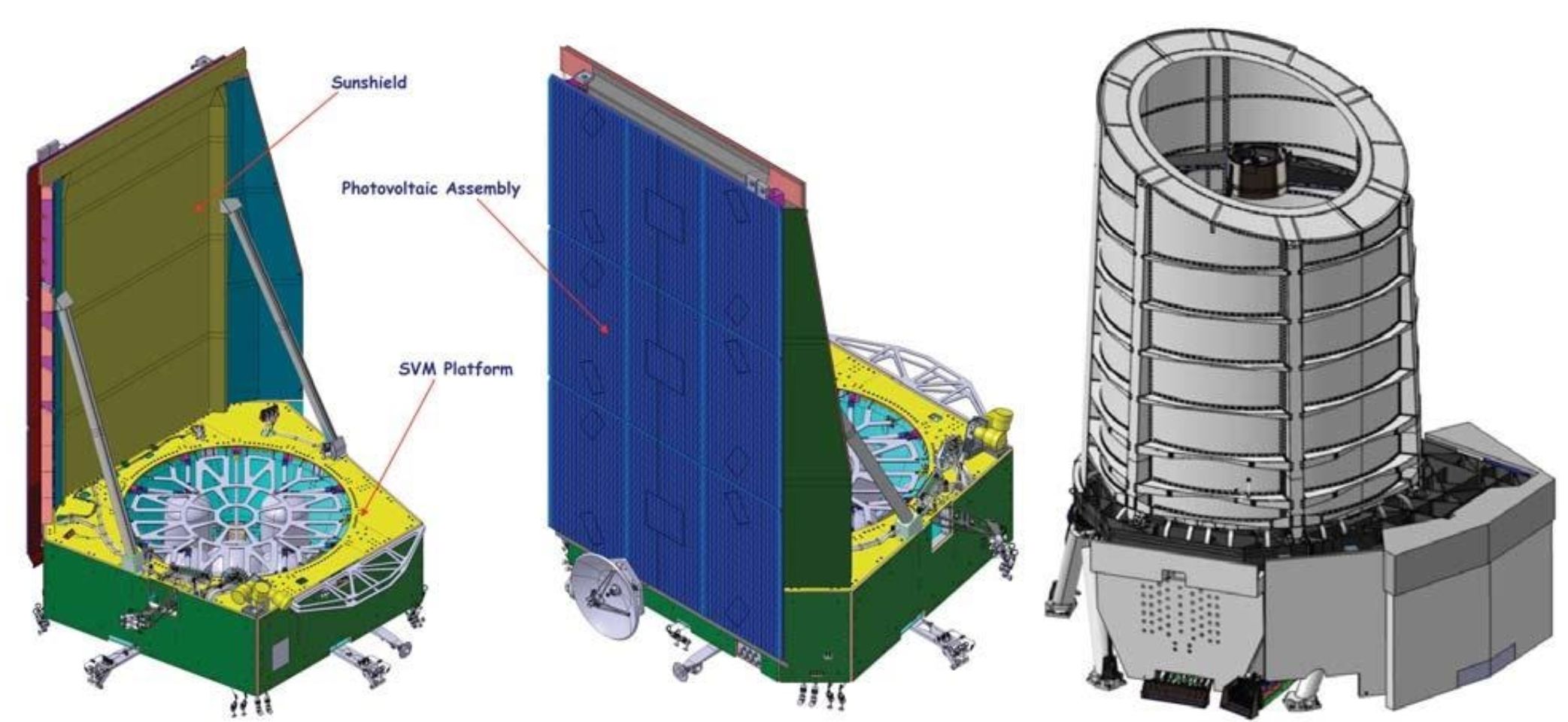}
    \caption{\Euclid spacecraft overview. From left to right: back view of the \ac{SVM}; front view of the \ac{SVM}; the integrated \ac{PLM} (not on the same scale as the \ac{SVM}). The sunshield is mounted to the \ac{SVM} structure and shields the \ac{PLM} structure from direct sunshine for all allowed spacecraft attitudes. The side of the \ac{PLM} exhibits three thermal radiators for the cooling of the \ac{VIS}, instrument electronics harness, and \ac{NISP}, from left to right, respectively. The telescope baffle is also used as thermal radiator. Three bipods are used to attach the \ac{PLM} onto the \ac{SVM} thereby providing maximum conductive thermal insulation.}
\label{appendix:FigCraft}
\end{figure}

In this paper, we consider three major satellite components: the service module (SVM) without sunshield, the payload module (PLM), and the sunshield (Fig. \ref{appendix:FigCraft}). By design, the sunshield is part of the \ac{SVM} and it is mounted to the \ac{SVM} with no mechanical contact with the \ac{PLM}. The \ac{PLM} contains the Korsch \ac{TMA} telescope and the two science instruments \ac{VIS} and \ac{NISP} as part of the optical train. All main \ac{PLM} components are mounted on the baseplate, which acts as a two-sided optical bench for the optical components and telescope baffle. Using a system of thermal radiators and heaters, including the telescope baffle that functions as a radiator, the \ac{PLM} instrument cavity is maintained at a required temperature of 135\,K. 

At the design level, many countermeasures were taken to isolate the \ac{PLM}, where the telescope and instruments are located, from the \ac{SVM}, where the ancillary subsystems reside and where most of the heat from the spacecraft is dissipated. The isolation was achieved by using high-efficiency thermal insulation and employing a set of three identical bipods, each of them composed of two glass fibre reinforced plastic struts, ensuring high stiffness to withstand the launch environment and low thermal conductance and thermal expansion.

The PLM telescope structure is based on an all-SiC concept; see \cite{EuclidSkyOverview}. Sintered Silicon Carbide offers special properties for the construction of a stable telescope (\citealp{Bougoin2012, Bougoin2018}). In particular its high steady state stability, i.e. the material ratio between the thermal conductivity and the coefficient of thermal expansion allows, in principle, the construction of a telescope that behaves homothetically under temperature changes. The high thermal conductivity (180\,W\,m\,\inv K\inv\, at room temperature) minimises the thermal gradients across the structure and the low coefficient of thermal expansion (CTE; 2.2 ppm\,K\inv\, at room temperature and 0.3 ppm\,K\inv at operational temperature) ensures that no optical aberrations occur for limited temperature changes. This isothermality is also enhanced by maximising thermal diffusivity (thermal conductivity over heat capacity) as is the case for SiC. In practice, the optical elements and some instrument parts are connected to the SiC structure by interposed elements made of Invar and glue, whose CTE is similar but not identical to that of SiC. This realisation together with a non-uniform temperature distribution introduces some residual optical aberrations, which are the subject of this study.

Another important element for stability is the control of the heat input into the PLM, which is isolated from the SVM by design. The two sources of power into the PLM instrument cavity are the VIS and NISP instruments. The VIS focal plane is mounted directly at the position of the telescope focal plane \citep{EuclidSkyVIS}. The heat dissipated in the focal plane read-out electronics is transferred to an external 1\,m$^2$ radiator, while the 36 CCDs are coupled to the SiC baseplate via thermal straps. The other VIS dissipating unit is the Readout Shutter Unit (RSU) which dumps its internal heat directly into the baseplate via radiation and conductance. NISP is instead a stand-alone instrument mounted on the baseplate with three bipods \citep{EuclidSkyNISP} mounted directly on the telescope baseplate. The average internal dissipation during the actuation of the filter and grism wheels is low and is damped directly to the baseplate through the bipods. The detection electronics instead requires high-efficiency thermal coupling to the baseplate, due to significant dissipation (4.8\,W) and low thermal gradient to be maintained with the baseplate; therefore, methane heat pipes are used. The NISP focal plane is the coldest part of the PLM, which must be kept below 100\,K. This is achieved by thermal coupling to the 2\,m$^2$ NISP radiator with 4 thermal straps. The heat dissipated by the instruments and its impact on the telescope stability is determined by the instrument operation cadence and the thermal design performance and is analysed in this paper.

The \ac{PLM} temperature control of the instrument cavity during science observations is carried out by setting the appropriate power of four heater groups in open loop to achieve the required temperatures at the specified instrument interface locations. The baseplate is not a controlled item, but its temperature is determined by the heater power settings and the environmental conditions of the spacecraft.

The 3\,$\sigma$ relative pointing error is required to be better than 25\,mas for a fiducial VIS exposure time of 700\,s. This is achieved with the fine guidance sensor (FGS) \citep{Racca2016} using a cold gas micro-propellant system (\ac{MPS}) for the actuation, which mainly provides a timely correction against the Solar pressure. Spacecraft vibrations are minimised by forcing the reaction wheels to stop spinning during exposures. These design measures ensure stable \ac{IQ} during an exposure. Longer term \ac{IQ} variations can only be caused by structural deformations due to changes in the thermal environment of the telescope or by degradation of the mirror surfaces. The latter effect is expected to occur on much longer timescales.

The thermal and structural design of the integrated spacecraft was verified by developing a \ac{STM} of the \Euclid satellite. This mechanical model of the spacecraft was partly assembled from the qualification models of the spacecraft components (\ac{SVM}, \ac{PLM}, sunshield, and instruments) and tested in 2019. The \ac{STM} underwent more extreme and risky thermal and structural testing conditions, which could not be applied to the flight model. The tests were adapted for the correlation with the mathematical models.

The \ac{PLM} flight model was assembled in \ac{ADS} in Toulouse and tested in the \ac{CSL} 2021, where it was placed under thermal vacuum to simulate the ambient conditions in orbit \citep{Gaspar2022}. The PLM thermal design with the aid of the thermal model was verified, and the in-orbit mathematical telescope model was derived from the model correlations obtained during this campaign. The thermal analysis performed before launch predicted an average $T_{\rm BP} = (131 \pm 4)$\,K, including $\pm3$\,K uncertainty. During the in-orbit phases examined in this paper the average $T_{\rm BP}$ has been $(131.9 \pm 0.3)$\,K. This achievement shows that the \ac{PLM} thermal control design performs extremely well and according to the predictions.

The \Euclid development consisted of a number of major common development milestones for the different system components. In this paper we refer to milestones for the spacecraft prime contractor: the \ac{PDR} held in Dec.\,2015, the \ac{CDR} held in Nov.\,2018, and the \ac{QR} held in Dec.\,2022. These milestones are accompanied by reviews where the requirements, implemented design, and expected performance are evaluated by independent panel members. 

After the launch of \Euclid on 1 July
2023, the plan was to devote the first month to the spacecraft commissioning phase followed by a 2-month performance verification phase before the start of the scientific programme. In practice various issues required the extension to 1.5 months of the spacecraft commissioning phase, followed by four months of performance verification phase that comprised also an extensive test campaign to verify the system software changes that were necessary to improve the pointing performance when observing sparse stars fields and in the presence of cosmic rays. This meant that scientific observations began only in December 2023. During the spacecraft commissioning phase, the telescope was put in focus by adjusting \ac{M2}. The \ac{IQ} parameters were verified during the performance verification phase.

\section{\label{appendix:sec:tools}Software applications to facilitate STOP analysis}

\subsection{\label{appendix:sub:MaREA-description}MaREA}

Two different approaches are available in MaREA for transient thermo-elastic analysis. In the direct approach, an interpolation function $F$(TMM, FEM) is created. At each time step, the temperatures are transferred from the TMM to the FEM, applying the $F$ function, and a NASTRAN simulation is performed to calculate the displacements. The second approach makes use of an analytical procedure called TEMASE (Temperature Mapping Sensitivity). A fixed number of unit-$\delta$T thermo-elastic load cases, equal to the number of thermal nodes, is run to create the so-called ``influence matrix''. A time sequence of displacements is obtained by multiplying the influence matrix by the matrix collecting all temperature time histories pertaining to the thermal nodes. The direct method is used once, to compute the elements of the influence matrix, while the matrix multiplication method is used to calculate the displacement time histories.
Synthesising, the process:
\begin{itemize}
    \item acquires temperature data from thermal analysis;
    \item establishes the correspondence between thermal and structural nodes and writes temperatures to the latter;
    \item loads each structural part with a unit-${\delta}T$, performs thermo-elastic analysis, and calculates the influence matrix elements;
    \item calculates the displacement time histories by matrix multiplication;
    \item transfers the computed displacements, runs the CODE V optical model, 
    and performs post-processing.
\end{itemize}
The most critical steps in the entire STOP process are the temperature interpolation from \ac{TMM} to \ac{FEM} and the transfer of displacements from FEM to CODE V. Both processes are executed automatically as part of the workflow. The temperature interpolation is based on an inverse distance weighting algorithm. To minimise the interpolation error, both TMM and FEM were split into several substructures, and the interpolation algorithm was applied to each one of them. After this step, each thermal node has a finite-element model counterpart.

\subsection{\label{appendix:sub:codeV} Using CODE V for the optical analysis}

\ac{IQ} variations are caused by the combination of the displacements of the optical surfaces attachment points and the thermal expansion of the curved mirrors. The first effect was introduced into the CODE V model of the telescope as output from the NASTRAN \ac{FEM} run under several thermal conditions, as explained in \cref{sec:design-phase}. For the second effect, a transfer function was defined, allowing us to derive the \ac{WFE} variations from the mean thermal variation of each optical surface. In synthesis, the definition of the transfer function is based on the \ac{FEM} of the telescope, including the detailed \ac{FEM} for optics and all SiC and Invar parts. A thermal load was applied to create a uniform temperature increase as large as 100\,K to get a significant change in order to avoid \ac{FEM} computation noise. The displacement of each optical surface and their deformations were integrated in the CODE V model. Once the global check was completed, the Zernike coefficients were calculated for all optical surfaces of interest. The resulting delta Zernike was then divided by 100 to get the sensitivity coefficient normalised to 1\,K. The second step was to add to the CODE V model the WFE variation caused by the mirror temperature changes. The combined WFE variation was obtained from the first 36 Zernike coefficients (using Fringe's indexing):
\begin{equation}
    {\rm d} \mathbf{z} = ( {\rm d}z_1, \dots, {\rm d}z_{36})\,.
\end{equation}
A quadratic model function of the Zernike variations was derived. This model was built by generating PSFs from telescopes with different WFE content using the CODE V model. The IQ metrics for each of these PSFs were computed and their variations as a function of Zernike variation were derived by a polynomial fitting. The outcome is one matrix of $36 \times 36$ coefficients per IQ metric. The variation of a given metric $M$ ($R^2$, $Q_{xx}$, $Q_{xy}$, $Q_{yy}$, and FWHM) is computed from
\begin{equation}
    {\rm d}M = \sum_{i=1}^{36} \sum_{j=1}^{36} a_{i,j}(M)\, (2\, z_i\, {\rm d}z_j + {\rm d}z_i\, {\rm d}z_j)\,,
\end{equation}
with $a_{i,j}$ the metric sensitivity coefficients, $z_i$ the baseline Zernike coefficients, computed with the CODE V model, and ${\rm d}z_i$ the variation of the Zernike coefficient. The final IQ metric is obtained from:
\begin{equation}
M=M_0+{\rm d}M\,,
\end{equation}
where $M_0$ is the metric baseline value, computed from the \ac{PSF} generated by the CODE V model. The assessment of ${\rm d}M$ remains accurate for low aberration. The error assessment of this method was evaluated with a Monte-Carlo analysis. 

\section{\label{appendix:sub:MPS-boom-redesign} Micro-propulsion boom redesign}

\begin{figure}
    \centering
    \includegraphics[width=0.5\columnwidth]{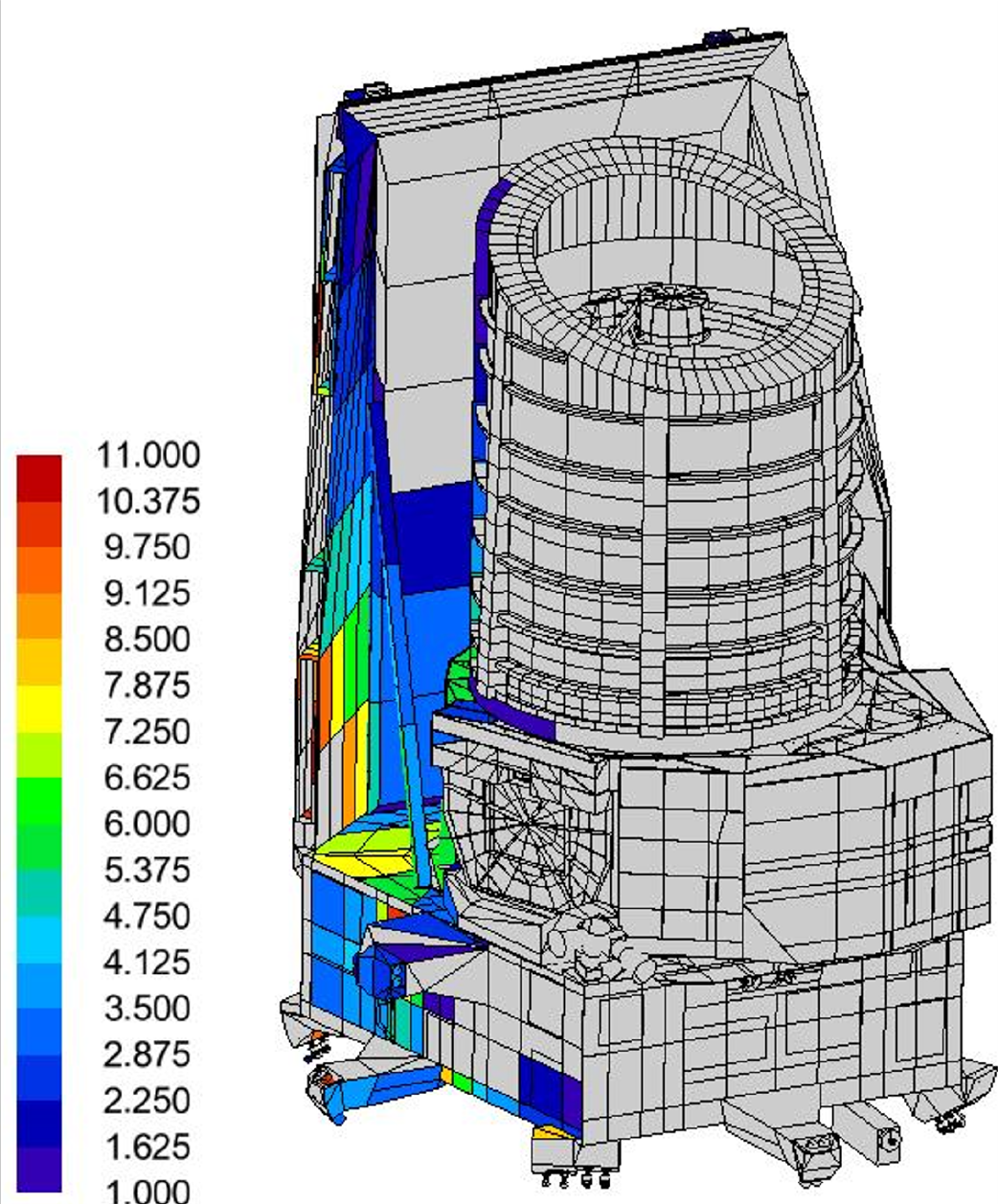}
    \caption{Temperature differences between Sun azimuth $-$8° and +8° at constant SAA. Grey areas are outside the colour code range. Modified MPS boom configuration.}
\label{appendix:FigTmodplot}
\end{figure}

Excluding the thrusters, the boom initial design was 470\,mm long and would be constantly exposed to the Sun. For \ac{SAA} $\leq$ 90\degree, the \ac{MLI} of the boom reached temperatures above 20\,\degree C and illuminated directly the \ac{VIS} radiator. The resulting maximum $T_\mathrm{BP}$ excursion was 1.4\,\degree C assuming \ac{EOL} conditions (see Fig.\,\ref{FigTBplot}a). Taking into account a 5\% change in ellipticity per degree of $T_{\rm BP}$, this would cause an ellipticity change of approximately 7\%, violating the \ac{IQ} requirement for ellipticity.

We reconfigured the \ac{MLI} on the topside of the boom such that it remained in shadow even at minimum \ac{SAA}. This caused the topside boom temperature to remain nearly constant when moving the Sun from the $-Y_{\rm SC}$ side to the  $+Y_{\rm SC}$ side. However, the sides of the boom facing  $+X_{\rm SC}$ experienced significant temperature variations. $T_\mathrm{BP}$ was found to be almost independent of \ac{SAA}, but continued to show a dependence on \ac{AA}. The maximum $T_{\rm BP}$ excursion was halved to 0.7\,\degree C still violating the ellipticity requirement. The Sun no longer illuminates the boom topside, but the sides facing the Sun reach temperatures of about 200\,\degree C. In the enclosure formed by the sunshield $+Y+Z$ wing, the \ac{PLM} side facing it, and the \ac{SVM} topside (all surfaces with a non-zero view factor to the Sun-facing side of the boom), the \ac{MLI} temperature increases by about 10\,\degree C, see \cref{appendix:FigTmodplot}. The \ac{PLM} baffle strip on the +Y side, unprotected by \ac{MLI}, undergoes a temperature raise of about 0.5\,\degree C. As the \ac{PLM} interior is directly coupled to the baffle, about the same effect is detected for the baseplate.

The next option was to reduce the length of the \ac{MPS} boom, the trade-off being the reduction of torque authority around the  $+Z_{\rm SC}$ axis. The minimum feasible length was determined to be 200\,mm. We still found dependencies on both \ac{AA} and \ac{SAA} (\cref{FigTBplot}b) but the maximum $T_\mathrm{BP}$ excursion at \ac{EOL} was further reduced to 0.6\,\degree C.
The simulation of the case without boom resulted in a maximum $T_\mathrm{BP}$ excursion of 0.5\,\degree C at \ac{EOL}. This showed that the half-length boom comes very close to cancelling the effects of the boom altogether. Based on this study, the design of the spacecraft was updated, incorporating the short boom.

\section{\label{appendix:IQ-perf-calculations} Time dependent IQ computations}

\begin{figure}
    \centering
    \includegraphics[width=0.8\columnwidth]{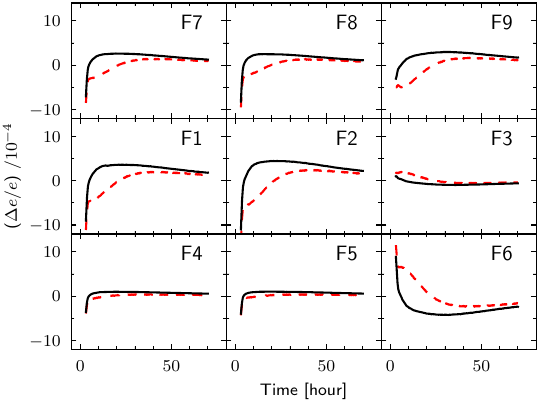}
    \caption{11\,000\,s variation of ellipticity $\delta e$ in the nine field points as function of time. Dashed lines: telescope only; solid lines: telescope and mirrors.}
\label{appendix:FigOpElf}
\end{figure}

\begin{figure}
    \centering
    \includegraphics[width=0.8\columnwidth]{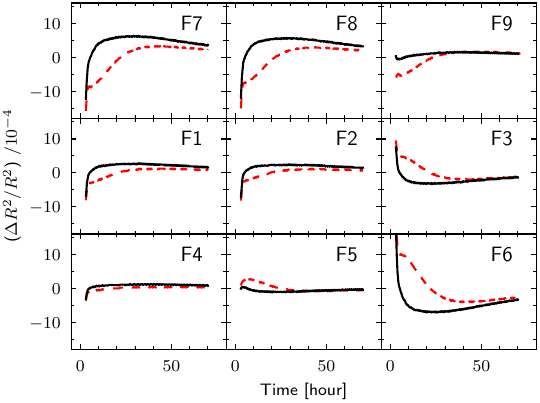}
    \caption{ 11\,000\,s variation of $\delta R^2/R^2$ in the nine field points as function of time. Dashed lines: telescope only; solid lines:  telescope and mirrors.}
\label{appendix:FigOpR2f}
\end{figure}

The 70\,h variation of the \ac{IQ} metrics in the field points is shown in Fig.\,\ref{appendix:FigOpElf} for the ellipticity and in Fig.\,\ref{appendix:FigOpR2f} for $R^2$. As explained in \cref{subsub:IQ-transient}, the long-term evolution driven by defocus and rapid initial variation driven by coma and astigmatism are the dominant components of the derived variations.

\section{\label{appendix:IQ-on-orbit}IQ determination from in-orbit VIS data}

To overcome the high glitch rate due to cosmic rays hitting the \ac{VIS} sensors, we decreased the standard deviation of the Gaussian weighting function from $\sigma =$ 0\farcs75 to 0\farcs25 when calculating the quadrupole moments for the derivation of the image quality parameters (see \cref{sec:requirements}) for stars in the VIS frames. This suppression of the wings of the \ac{PSF} causes a systematically smaller value of $R^2$. The \ac{IQ} values we derived were still sensitive to environmental changes, so we can use them as diagnostic for changes in \ac{IQ}. In our analysis, we removed VIS frames that were affected by strong solar flares using a merit function derived from $e_1$, $e_2$, and $R^2$ covariances. However, we found that some areas in the FPA are more affected by solar X-rays than others because of selective shielding by the sunshield \citep{laureijs2024}. For the value of $R^2$ averaged over the FPA we therefore took the weighted mean per \ac{FPA} area. This weighting was not performed to obtain the average value of ellipticity components due to the large intrinsic variation of the ellipticity (a factor of approximately 5) over the FPA.

The large width of the VIS band causes a strong PSF dependency on the colour of the stars in the field. From the inspection of individual VIS frames we estimate a $\approx$4\% increase of $R^2$ per magnitude of {\it Gaia} $B_{\rm p}R_{\rm p}$. Furthermore, the PSF is affected by the brighter-fatter effect, inherent in the usage of CCD sensors. To minimise these effects we only considered stars that were also detected by {\it Gaia} within the colour range $1.5< B_pR_p < 2.5$ providing approximately flat spectral energy distribution in the VIS band, and with a {\it Gaia} magnitude in the range $18.9 < m_G < 19.9$ to constrain \ac{IQ} variations due to stellar brightness dependencies. The PSF measurements were binned in 9 equal areas on the VIS FPA to investigate the in-orbit \ac{IQ} variations over the focal plane.

\end{appendix}
\end{document}